# In situ elucidation of mechanisms governing crack transition to plasticity arrest

Abdalrhaman Koko [1,2]*, Bemin Sheen [3], Caitlin Green [1], and Fionn Dunne [2]

[1] National Physical Laboratory, Hampton Road, Teddington TW11 0LW, United Kingdom

[2] Department of Materials, Royal School of Mines, Imperial College London, UK

[3] Department of Mechanical Engineering, Imperial College London, UK

* Corresponding author. E-mail: abdo.koko@npl.co.uk

## Abstract

Despite extensive theoretical treatment of short- to long-crack transitions, direct experimental quantification of how elastic and plastic energy contributions evolve at the crack tip during arrest has remained absent. In this study, we present an in situ investigation of crack propagation in cold-worked AA-5052 using high-resolution scanning electron microscopy digital image correlation (SEM-DIC) and electron backscatter diffraction (EBSD). By reconstructing local crack-tip fields from measured displacement data, we extract mode I and II stress intensity factors and both elastic and elastoplastic energy release rates ($\Delta J_E$ and $\Delta J_p$). The results show that microstructure-sensitive cracks propagate in a mixed-mode manner at low driving force and transition to plasticity-dominated and load-aligned crack, arrested as the crack-tip process zone develops and expands multiple grains. This transition is identified through the divergence of elastic and elastoplastic energy measures ($\Delta J_E \geq \Delta J_P$), crack-tip blunting, slip-band emission, and the emergence of localised plastic deformation. These findings demonstrate that crack arrest coincides with a measurable transition in crack-tip energy partitioning and with process-zone expansion beyond grain-scale dimensions. The results establish an experimentally measurable energy-~~based~~partition criterion for crack arrest and demonstrate that fracture regime transition is governed by process-zone expansion relative to microstructural length scales rather than crack length.

# 1. Introduction

Crack propagation in structural metals remains a key concern in high-performance applications such as aerospace, marine, and transportation systems. Ensuring structural integrity over time depends not only on delaying crack initiation but also on understanding how cracks grow and, critically, how they might arrest [1,2]. Yet, in high-ductility alloys, such as cold-worked aluminium, crack behaviour is especially complex due to the interplay between residual plastic deformation and heterogeneous microstructures [3,4]. While long cracks are generally well-described by global fracture mechanics parameters, what is described in the literature as *short cracks* are not solely controlled by far-field applied stresses, but are substantially affected by interactions with the local microstructure [5–7], such as grain boundaries [8,9], intragranular misorientations [8,10], and residual stress distributions [11,12]. These interactions create highly variable crack growth rates and tortuous crack paths as the crack advances through individual grains [9,13,14]. But these distinctions between microstructure-sensitive cracks do not simply scale down from long crack behaviour [15,16], and the typical classification as either “short” or “long,” with the implicit assumption that length controls, does not convey whether fracture is governed by microstructural interactions or by continuum-level driving forces.

This distinction gets muddier when predicting the transition from microstructure-sensitive crack growth to arrest in ductile materials, which is essential for advancing damage-tolerant design strategies, but despite decades of fracture research, a clear and continuous understanding of how cracks transition from microstructurally sensitive paths to plasticity-dominated propagation has remained elusive, especially [17]. While ductile tearing is commonly accepted as the primary crack propagation mechanism [18–20], the mechanisms of microstructure-sensitive crack propagation and their transition to ductile tearing require further investigation. Most available studies either focus on long crack growth using global parameters such as $K_{IC}$ or $J_{IC}$ [21–23], or they examine microstructural crack-microstructure interactions in isolation [4,24–27], either through post-mortem analysis [4,28,29], modelling [30–32], or discontinuous observations [33,34], often limited by spatial resolution or lacking full-field strain data [35–37]. While prior studies have addressed aspects of microstructure-sensitive crack growth and ductile tearing separately, and ~~heterogeneities, becomes a~~proposed that crack-tip plastic-zone development governs the transition from microstructure-sensitive to ductile crack behaviour, direct in situ quantification of crack-tip driving forces during this transition remains missing. This transition is crucial to both the scientific understanding of crack evolution and the engineering prediction of failure, particularly in FCC alloys where crack blunting and plastic shielding compete with surface formation.

In this study, we directly address this missing link. Using in situ scanning electron microscopy-based digital image correlation (SEM-DIC) combined with electron backscatter diffraction (EBSD), we follow crack propagation in a cold-worked 5052 aluminium alloy (AA-5052) as the crack-tip process zone evolves from being confined within individual grains to spanning multiple grains. This approach provides the experimental evidence that crack behaviour is governed not by crack length, but by the relative size of the process-zone.

The remainder of this paper is structured as follows. Section 2 describes the material, in situ SEM-DIC methodology, and displacement-driven reconstruction framework. Section 3 presents crack evolution and reconstructed fracture metrics. Section 4 discusses the transition from microstructure-sensitive propagation to plasticity-dominated arrest in terms of energy partitioning and process-zone development. Finally, the key findings and implications are summarised.

## 2. Methodology

This section describes the material, experimental configuration, and the displacement-driven reconstruction framework used to extract crack-tip driving forces. The methodology integrates in situ SEM-DIC measurements with EBSD characterisation and finite element reconstruction. The experimental components define the microstructural state and kinematic fields, while the numerical framework reconstructs stress and energy measures consistent with the measured displacements.

### ~~1.1.~~2.1. Material and pre-mechanical testing characterisation

An ASTM E647 compact tension (CT) specimen was fabricated from an aerospace-grade AA-5052 aluminium alloy plate with a rolling texture (see Fig. S3) and an average grain diameter of 31.7 ± 8.0 µm. The CT specimen's notch was aligned with the rolling direction. AA-5052 was selected as a model system due to its strain-hardened microstructure, high work-hardening rate, and relevance to ductile fracture in non-heat-treatable aluminium alloys [38–40]. These features make it ideally suited for studying plasticity-driven crack transition mechanisms in a controlled, well-characterised environment.

The chemical composition of the AA-5052 plate was confirmed using a HITACHI X-MET8000 Expert CG X-ray fluorescence (XRF) spectrometer to perform positive material identification. The detailed elemental composition results are presented in Table 1. In addition, the plate residual stresses were subsequently measured using the cos α method via Pulstec μ-X360 diffractometer with Cr-Kα radiation and a 0.3 mm pinhole collimator, achieving a spot size of approximately 1 mm, from a depth of around 10 μm at a 25° inclination.

Table 1: Measured chemical composition of the AA-5052 plate in weight%.

| Al | Mg | Fe | Cr | Si | Mn | Cu | Zn |
|---|---|---|---|---|---|---|---|
| 96.047 ± 0.255 | 3.137 ± 0.176 | 0.360 ± 0.009 | 0.223 ± 0.037 | 0.133 ± 0.020 | 0.050 ± 0.004 | 0.043 ± 0.004 | 0.007 ± 0.003 |

The specimen was prepared using precision wire electrical discharge machining to ensure dimensional accuracy and to minimise residual stresses introduced during the cutting process. The final geometry of the CT specimen featured a thickness ($B$) of 2.82 ± 0.02 mm, an effective width ($W$) of 13 ± 0.08 mm, notch length of 6.5 mm, and radius of 0.19 mm, with the pin diameter machined to 3.25 mm, as illustrated in **Fig. 1**a. The specimen was fatigue pre-cracked at a frequency of 10 Hz, with a load ratio (R) of 0.1, until a sharp crack of 76.42 ± 0.70 μm was achieved, as shown in **Fig. 1**c.

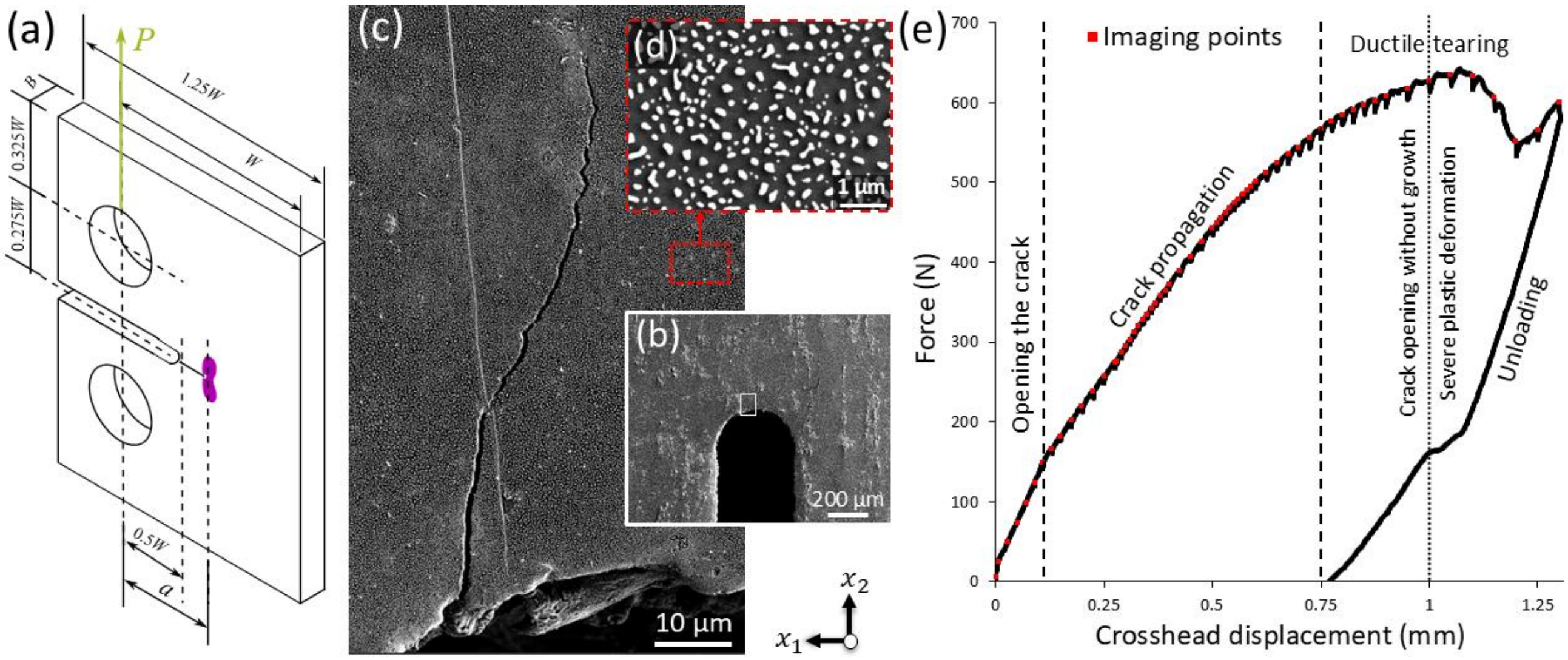

**Fig. 1. Imaging and mechanical testing of the specimen.** (a) Schematic of the ASTM E647 CT specimen geometry, showing key dimensions including specimen thickness ($B$), effective width ($W$), and notch configuration. The loading ($P$) direction is indicated by green arrows, with the other side fixed. (b) Low-magnification SEM image of the CT notch and pre-crack region. (c) Higher-magnification ETD-SEM image of the fatigue-induced crack tip area (white box in (b)). (d) ETD-SEM image of the gold nanoparticle speckle pattern used for DIC, with an average particle size of 225 ± 16 nm. (e) The in situ SEM test's nominal load-displacement curve is segmented as observed via the SEM images. Red markers indicate the 61 imaging points where SEM snapshots were captured to correlate crack evolution with loading conditions.

Following fatigue pre-cracking, the specimen was mechanically mirror-polished to remove any surface deformation, and the microstructure around the notch was mapped using EBSD performed with a Zeiss Auriga 60 SEM equipped with an Oxford Symmetry 2 EBSD detector. The specimen was positioned at a working distance (WD) of 14 mm from both the SEM pole piece and the EBSD detector. An electron beam with an accelerating voltage of 20 keV and a 120 µm aperture (yielding a beam current of approximately 10 nA) was employed for diffraction data collection at a step size of 0.5 µm. EBSD scans performed after polishing and before monotonic loading confirmed that the microstructure near the fatigue pre-crack tip remained consistent with the cold-worked state, with no signs of residual plasticity or substructure development (Fig. 3a).

For HR-EBSD analysis [41] of the sub-microstructure of the cold-rolled AA-5052, high-resolution 1244 x 1024-pixel Kikuchi patterns were collected with a 100-millisecond exposure per pattern and 50 nm step size. Subsequent EBSD analyses – including slip trace analysis – were performed using MTEX [42], and HR-EBSD analyses were performed using CrossCourt

v4, with the patterns segmented into 40 regions before cross-correlation with a reference pattern, selected using the method described in [43], to estimate the local distribution of residual stresses and geometrically necessary dislocation (GND) density [44]. In addition to the EBSD analysis, the chemical composition of precipitates present in the microstructure was investigated using energy dispersive X-ray spectrometry (EDX) via an Oxford X-Max80 detector at an accelerating voltage of 10 kV and a WD of 5 mm.

Sequential EBSD mapping during crack propagation was not performed because the gold speckle pattern required for SEM-DIC precludes reliable diffraction acquisition. In addition, following the transition to plasticity-dominated behaviour, the high level of surface deformation resulted in substantial degradation of Kikuchi pattern quality, preventing quantitative extraction of post-deformation GND fields near the crack tip. Consequently, EBSD-derived metrics reported here characterise the initial microstructural state rather than dynamic crack-tip evolution. Additionally, serial sectioning 3D EBSD was attempted in the arrested crack region. However, the severe plastic deformation and lattice distortion within approximately 70–80 μm of the crack tip led to substantial degradation of Kikuchi pattern quality, preventing reliable orientation indexing in the immediate crack-front vicinity (see Fig. S8). As a result, quantitative GND evolution at the crack tip could not be extracted post-deformation.

## ~~1.2.~~2.2. Mechanical testing

The pre-cracked specimen was subjected to monotonic tensile loading within an Apreo S2 SEM to facilitate further propagation of the initial fatigue crack. The tensile test was conducted using a 5 kN Deben in situ tensile stage at a constant displacement rate of 0.1 mm/min. The specimen was positioned at a WD of 13.6 mm from the SEM pole piece, and imaging was performed using an electron beam at 5 kV and 1 nA.

The specimen was continuously monitored during the test, but the test was periodically paused under displacement control for 2 minutes to capture high-resolution images of the crack growth (red dots in **Fig. 1**~~E~~e), using an Everhart-Thornley Detector (ETD) with a pixel size of 25.8 nm. The contrast between the high-density gold speckles, which appeared as bright spots (as shown in **Fig. 1**~~D~~d), and the AA-5052 matrix was leveraged to facilitate subsequent DIC analysis for monitoring crack propagation.

Following the mechanical testing, the specimen was scanned using an Alicona InfiniteFocus optical microscope to generate a three-dimensional dataset of the region surrounding the notch. The scanning was performed with a 50x objective lens using an Alicona G5 model and processed using Alicona MeasureSuite software v.5.3.6.

### ~~1.3.~~2.3. Digital image correlation analysis

For SEM-DIC, a gold speckle pattern was applied to the specimen surface for high-resolution crack growth monitoring using SEM-DIC. A uniform 10 nm thick gold film was initially deposited onto the polished specimen using a Quorum Q150T coater at an accelerating voltage of 20 kV for 10 seconds. To achieve an optimal speckle pattern suitable for SEM-DIC analysis, a water vapour-assisted gold remodelling method [45,46] was employed for 6 hours at 250 °C, facilitating the formation of discrete gold speckles with an average diameter of 225 ± 16 nm. This speckle size provided the necessary contrast and resolution to accurately monitor crack propagation during subsequent mechanical testing (Fig. 4).

Heating the specimen at 250 °C for 6 hours, as part of the gold remodelling process, lies within the stress-relief annealing range for AA-5052 [47,48], which promotes dislocation recovery without recrystallisation, further reducing any residual stress introduced by the fatigue pre-crack.

DIC was employed to analyse the SEM images acquired during the in situ tensile testing. Before the DIC analysis, the images were aligned to a common field of view to eliminate rigid body movement, using the linear stack alignment plugin in Fiji ImageJ, leveraging the Scale-Invariant Feature Transform (SIFT) algorithm.

DIC data processing was conducted using DaVis v.10.2 software (LaVision GmbH). The analysis followed the best practices outlined in the iDICs guidelines [49]. Key parameters for the DIC process included a subset size of 17 pixels and a step size of 11 pixels, balancing spatial resolution and precision [50], and resulting in a displacement field with a 286 nm step size. The subset shape function was set to second order, enabling the capture of non-linear deformation fields around the crack tip. For image matching, the Zero-Normalised Sum of Squared Differences (ZNSSD) criterion was applied, offering robustness against intensity fluctuations and image noise, thereby enhancing correlation accuracy. A bicubic spline interpolant was used to generate smooth and continuous displacement fields, critical for precise strain mapping in regions of high deformation gradients. In addition, to correct for any residual rigid body movements, the method described in [51] was implemented. This approach involved identifying the point with the absolute minimum displacement within the image set and designating it as the origin.

### ~~1.4.~~2.4. *J*-integral and stress intensity factors calculations

We adopt a domain *J*-integral ($\Delta J$) formulation, computed between successive crack tip configurations based on full-field experimental displacement data from SEM-DIC. This avoids the need for stationary crack tip assumptions and instead directly quantifies the energy release and stress intensity factors (SIFs) during discrete crack advances. This approach is consistent with previous studies [52] that have used domain-integral methods with experimentally measured full-field displacement, including DIC [53,54], stereo-DIC [55,56], and even digital volume correlation [57,58].

Using a custom-made MATLAB script [59], a finite element (FE) model was created in Abaqus v.6.14 from the SEM-DIC field of view, excluding the crack geometry, to calculate the strain energy release rate and the mixed-mode SIFs at the crack tip, as shown in **Fig. 2**~~A~~a and b. The model consisted of a rectangular grid composed of 4-node plane stress elements (Abaqus CPS4), designed to match the DIC grid, with no interpolation, ensuring a direct correlation between experimental and numerical data. Then, the experimentally measured DIC displacement field was applied as a boundary condition at each node in the FE model [59]. Similarly, after registration of the EBSD data in MTEX, each element of the model's mesh is assigned an orientation by interpolating the orientations EBSD map at the location of each element centroid through inverse distance spatial interpolation.

In essence, the finite element framework is employed in an inverse manner: experimentally measured displacement fields are imposed as nodal boundary conditions, and the model reconstructs stress and energy fields consistent with these kinematics. It is not used to predict crack growth.

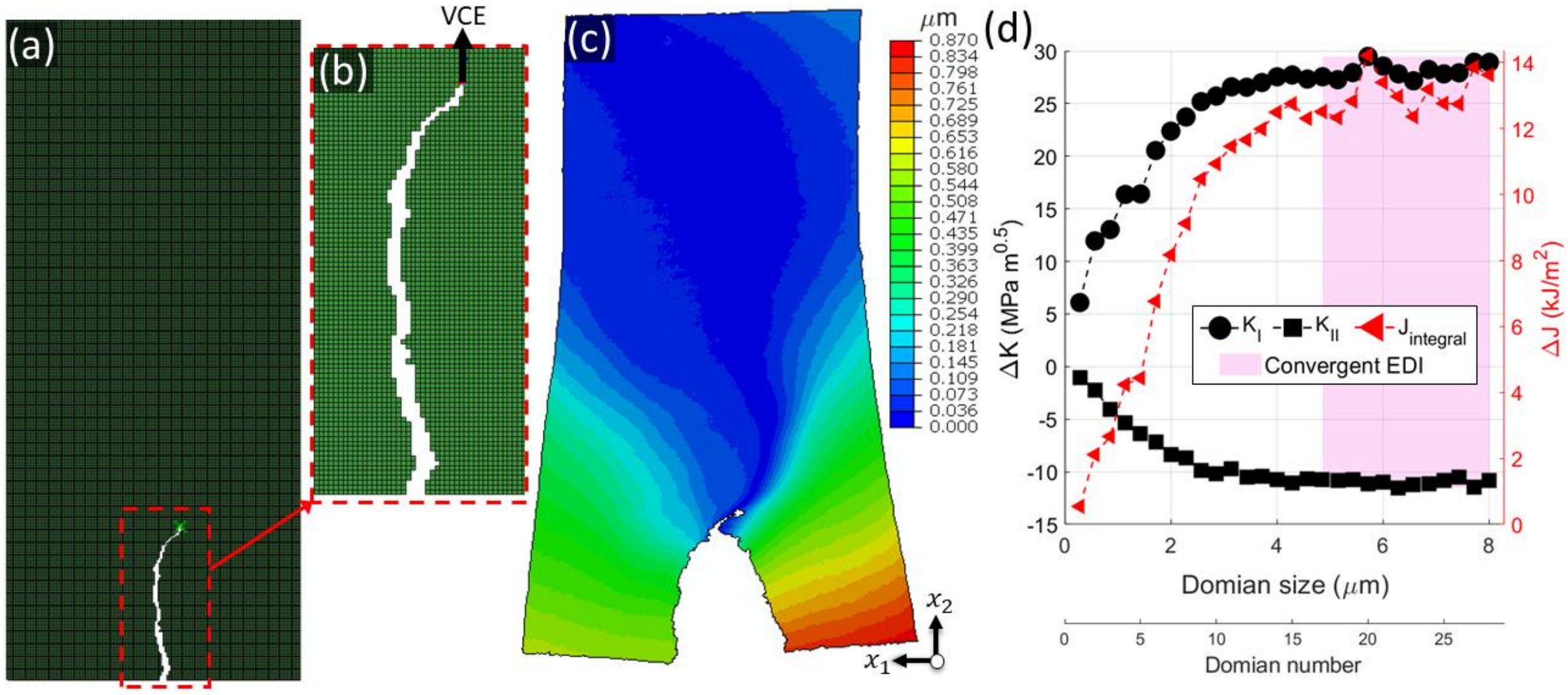

**Fig. 2. Extraction of DIC-based SIFs.** (a) A finite element mesh of the region of interest was created in Abaqus, replicating the experimental DIC field of view, with the physical crack geometry excluded from the mesh. (b) Zoomed view of the crack path, showing mesh refinement around the crack flanks and the virtual crack extension (VCE) direction used for *J*-integral evaluation. (c) Displacement magnitude in the deformed configuration after node-wise experimental DIC data were applied at 0.32 mm crosshead displacement. (d) Domain integral results for mode I ($\Delta K_I$), mode II ($\Delta K_{II}$), and the *J*-integral at 0.9 mm crosshead displacement. The shaded pink region marks the domain range used for extracting converged fracture parameters.

To calculate the elastic part of the strain energy release rate ($\Delta J_E$), we used AA-5052 anisotropic stiffness ($C_{11} = 106.75, C_{44} = 28.34, C_{12} = 60.41$ in GPa [60]) after it was transformed from the reference crystal coordinate system to the crystal orientation (or frame of reference) using a matrix ($\mathbf{g}$) constructed from the EBSD-determined Bunge-Euler angles ($\phi_1, \Phi, \phi_2$) [61], as shown in question (1) to (4).

$$\mathbf{g}_Z(\phi_1) = \begin{bmatrix} \cos(\phi_1) & \sin(\phi_1) & 0 \\ -\sin(\phi_1) & \cos(\phi_1) & 0 \\ 0 & 0 & 1 \end{bmatrix}, \tag{1}$$

$$\mathbf{g}_{X'}(\Phi) = \begin{bmatrix} 1 & 0 & 0 \\ 0 & \cos(\Phi) & \sin(\Phi) \\ 0 & -\sin(\Phi) & \cos(\Phi) \end{bmatrix}, \tag{2}$$

$$\mathbf{g}_{Z''}(\phi_2) = \begin{bmatrix} \cos(\phi_2) & \sin(\phi_2) & 0 \\ -\sin(\phi_2) & \cos(\phi_2) & 0 \\ 0 & 0 & 1 \end{bmatrix}, \tag{3}$$

$$\mathbf{g}(\phi_1, \Phi, \phi_2) = \mathbf{g}_{Z''}(\phi_2)\, \mathbf{g}_{X'}(\Phi)\, \mathbf{g}_Z(\phi_1). \tag{4}$$

For an active rotation of the crystal axes in the sample reference frame, the rotation matrix (R) is used, where $\mathbf{R} = \mathbf{g}^T$. The rotated stiffness matrix ($\mathbf{C}'$) is then produced by rotating $\mathbf{C}$ as a 4th rank tensor, as expressed in equation (5). After rotation, $\mathbf{C}'$ is converted to a 6x6 2nd order tensor with Abaqus' ordering of the entries (leading diagonal: $C_{11}$, $C_{22}$, $C_{33}$, $C_{12}$, $C_{13}$, $C_{23}$).

$$\mathbf{C}' = \mathbf{R} \cdot \mathbf{R} \cdot \mathbf{C} \cdot \mathbf{R}^T \cdot \mathbf{R}^T, \tag{5}$$

$$C'_{ijkl} = R_{im} R_{jn} R_{kp} R_{lq} C_{mnpq}. \tag{6}$$

To account for nonlinearity and plastic deformation near the crack tip after the crack stopped growing, the full (elastoplastic) strain energy release rate ($\Delta J_p$) was calculated using the crystal plasticity framework as implemented in [62], briefly explained here. The mechanical part of the deformation gradient ($\mathbf{F}$) is decomposed multiplicatively into the elastic and plastic deformation gradients ($\mathbf{F}_e$ and $\mathbf{F}_p$ in equation (7)).

$$\mathbf{F} = \mathbf{F}_e \ \mathbf{F}_p. \tag{7}$$

The plastic velocity gradient ($\mathbf{L}_p$) is related to $\mathbf{F}_p$ and the rate of change of the plastic deformation gradient ($\dot{\mathbf{F}}_p$) as shown in (8). $\dot{\mathbf{F}}_p$ is expressed in terms of $\mathbf{F}_p$ at the start of a time increment, i.e., $\left(\mathbf{F}_p\right)_t$. Rearranging to equation allows $\mathbf{F}_p$ to be updated, based on $\mathbf{L}_p$, at the end of each increment.

$$\mathbf{L}_p = \dot{\mathbf{F}}_p \mathbf{F}_p^{-1} = \frac{\mathbf{F}_p - \left(\mathbf{F}_p\right)_t}{\Delta t} \mathbf{F}_p^{-1}, \tag{8}$$

$$\mathbf{F}_p = \left(\mathbf{I} - \mathbf{L}_p \Delta t\right)^{-1} \left(\mathbf{F}_p\right)_t, \tag{9}$$

where $\mathbf{L}_p$ is then obtained from the summation of shear rates on slip systems with slip direction ($\boldsymbol{s}$) and slip normal ($\boldsymbol{n}$). The slip rate on the $\alpha^{\text{th}}$ slip system, $\dot{\gamma}^a$ is calculated via a slip law, such as the slip law based on the Orowan slip equation, which is provided in equation (11) [63]:

$$\mathbf{L}_p = \sum_a \dot{\gamma}^a s^a \otimes n^a, \tag{10}$$

$$\dot{\gamma}^{a} = \rho_{\mathrm{m}} \nu (b^{a})^{2} \exp\left(-\frac{\Delta F^{a}}{kT}\right) \sinh\left(\frac{\Delta V^{a}}{kT}(\tau^{a} - \tau_{c}^{a})\right), \tag{11}$$

where $k$ is the Boltzmann constant, $T$ is the temperature, and $\tau$ is the resolved shear stress. Other constants are listed in Table 2 and selected based on published data for AA-5052 [65], and are consistent with independently measured monotonic tensile behaviour for the present material batch. The activation volume is high to eliminate rate-dependent effects [64]. Because the displacement field is measured experimentally at each increment, the reconstructed stress and energy fields are constrained by kinematic compatibility, reducing dependence on constitutive extrapolation. The incremental *J*-integral formulation avoids assumptions of stationary crack-tip fields and allows direct evaluation of energy evolution during discrete crack advances.

Table 2: AA-5052 properties used in equation (11) [65].

| Property | Symbol | Value | Units |
|---|---|---|---|
| Density of mobile dislocations | $\rho_m$ | $10^{12}$ | $m^{-2}$ |
| Frequency of dislocation jumps | $\nu$ | $10^{11}$ | $s^{-1}$ |
| Magnitude of the Burgers vector | b | 0.286 | nm |
| Critical resolved shear stress | $\tau_c$ | 65.74 | MPa |
| Activation energy | $\Delta F$ | 0.75 | eV |
| Activation volume | $\Delta V$ | $400b^3$ | $m^3$ |

To set the crystal orientation in the UMATs, the Euler angles in the Bunge convention of each material definition is required to be specified in its properties. As such, a material is defined for each element to model each orientation. It has been observed that more accurate results have been achieved by using a 3D mesh with UMATs compared to a 2D mesh. Accordingly, the model builds a mesh of C3D8 elements when using UMATs. Validation of this model can be found in the Supplementary Materials.

The elastic strain energy release rate ($\Delta J_E$) enabled the calculation of mode I (opening mode) and mode II (in-plane shear mode) SIFs at the crack tip using the interaction integral method implemented in Abaqus [66], which deconvolutes the forces acting on the crack. Plus, this dual approach allows us to identify the onset of plasticity-driven crack arrest by comparing $\Delta J_E$ and $\Delta J_P$. For $\Delta J_E$ and $\Delta J_P$, the equivalent domain integral (EDI) method [67] was coupled with the virtual crack extension (VCE) technique [68] to calculate the strain energy release rate, or the $J$-integral, per domain as the domain expanded from the crack tip in the VCE direction at 286 nm intervals. The VCE was set in the average direction that the crack took in the next interval, up to where the crack stopped growing, after which the crack was assumed to be parallel to the $x_2$-axis.

To further validate our approach, we examine the convergence behaviour of the domain integral with increasing domain size. As shown in the pink shaded area in **Fig. 2**d, the $J$-integral and stress intensity factors reach stable values over a well-defined domain region. This convergence confirms that the process zone is fully captured and that local unloading or DIC noise artefacts do not unduly affect the energy release rate. Initial non-convergence was attributed to the presence of highly localised plasticity [69,70] and insufficient strain data discretisation near the crack tip [71]. Variance within this region served as an indicator of convergence stability. Also, the sign of the mode II SIF is irrelevant, as it depends on the specific nodal arrangement at the crack tip and does not have a physical interpretation [72].

Comparatively, the CT specimen geometry and the crack length ($L$) was used to calculate the ASTM E1820 stress intensity factor ($K_{ASTM}$) per equation (12) [73], with its uncertainty calculated via the Monte Carlo method.

$$K_{ASTM} = \frac{P}{BW^{1/2}}\frac{2+a/W}{(1-a/w)^{3/2}}\left[0.886+4.64\left(\frac{a}{W}\right)-13.32\left(\frac{a}{W}\right)^2+14.72\left(\frac{a}{W}\right)^3 - 5.6\left(\frac{a}{W}\right)^4\right], \quad (12)$$

$$a = 0.5W + L.$$

## ~~2.~~3. Results

This section presents the evolution of crack morphology, strain fields, and reconstructed fracture metrics during monotonic loading. We first describe microstructure-sensitive crack propagation, followed by the transition to crack arrest and plasticity-dominated deformation.

### ~~2.1.~~3.1. Crack opening and propagation

As shown in **Fig. 1**e, the nominal load-displacement curve was divided into key regimes, as observed from the SEM images: initial crack opening, stable crack propagation, crack opening without growth, severe plastic deformation, and unloading. Further, the stable crack propagation regime was segmented with respect to its crystallographic orientation and grain boundary characteristics, as summarised in Table 3 and shown in **Fig. 3**a. Overall, crack propagation was intragranular, starting in Grain A and terminating within Grain B, spanning 11 segments. Note that the word *intragranular* here is relative to our definition of grain boundaries at a 12° misorientation threshold.

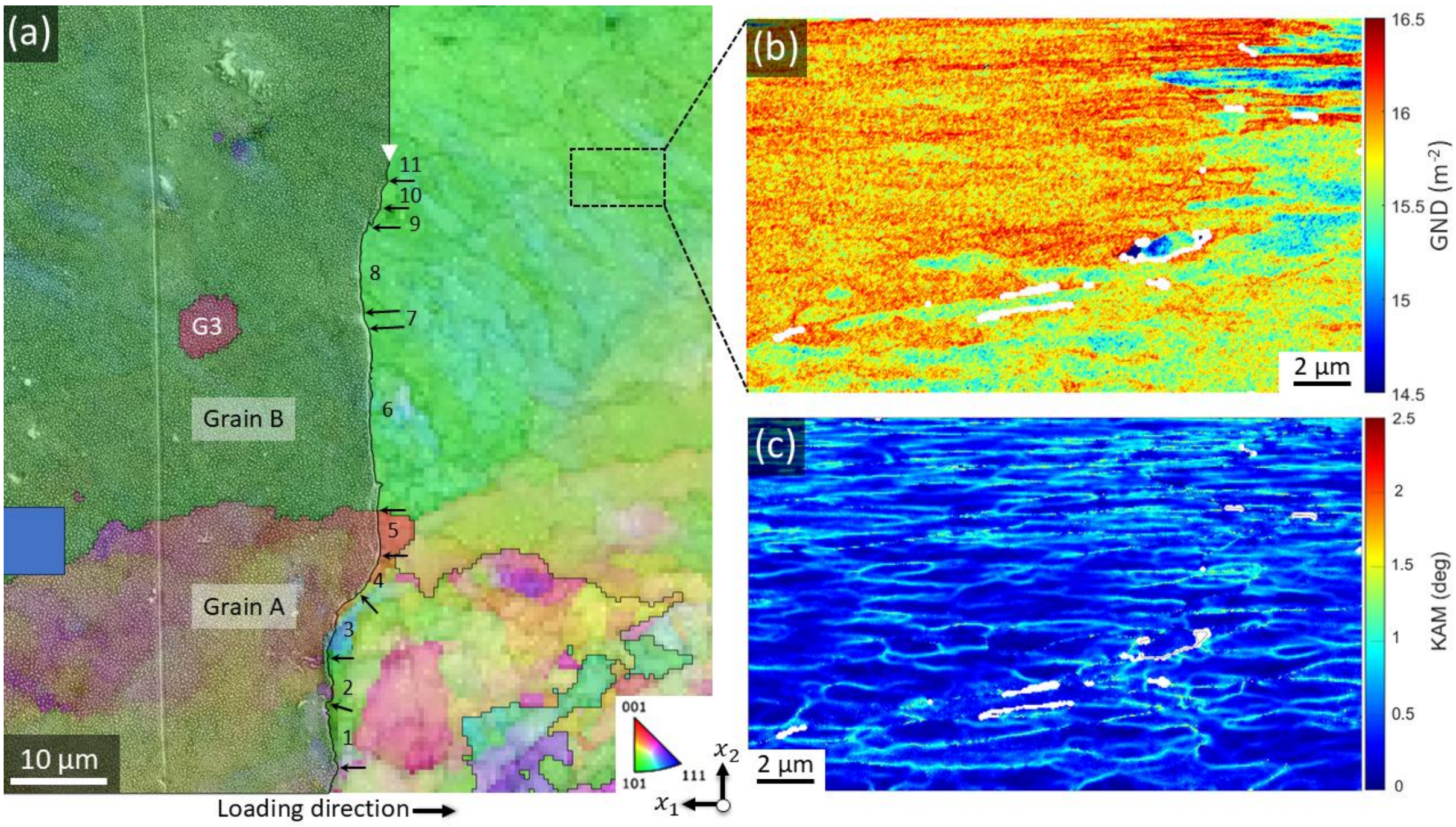

**Fig. 3. AA-5052 microstructure**. (a) EBSD orientation map acquired before testing using a 0.5 µm step size, with the stationary portion of the crack, recorded at 0.75 mm crosshead displacement, superimposed. The final crack path (taken from **Fig. 4**a) is superimposed here to illustrate how the crack interacted with the initial microstructure prior to deformation. Grains A and B are determined based on a 12° grain boundary misorientation threshold; using a lower threshold reveals subgrain structures indicative of cold work. (b) Geometrically necessary dislocation (GND) density map derived from high-resolution EBSD data acquired at a 50 nm step size. (c) Kernel average misorientation (KAM) map calculated using a 3-pixel window for the same field of view as B, highlighting local plastic strain gradients. The maps in B/C were collected away from the crack, as a general representative of the initial microstructure.

Table 3. Crystallographic tracing of the crack. Characteristics of the crack path in the AA-5052 as traced from SEM images and the EBSD map (**Fig. 3**a and **Fig. 4**a). Sections are shown in **Fig. 3**a. Crosshead displacement, in mm, and force, in N, tabulated below, are taken at the end of the segment.

| Grain | no. | mm, N | Length (µm) | Crack plane |
|---|---|---|---|---|
| A | 1 | 0.09, 124 | 0 (7.18 ± 0.09) | $(1\bar{1}\bar{1})$ ± 3.05° |
| | 2 | 0.15,183 | 4.61 ± 0.30 | - |
| | 3 | 0.25, 258 | 8.10 ± 0.01 | - |
| | 4 | 0.32, 312 | 4.59 ± 0.19 | - |
| | 5 | 0.35, 335 | 5.34 ± 0.02 | $(\bar{1}\bar{2}1)$ ± 3.30° |
| Σ13 ± 2.76° grain boundary | | | | |
| B | 6 | 0.4, 372 | 18.83 ± 0.28 | $(112)$ ± 1.44° |
| | 7 | 0.45, 408 | 2.34 ± 0.16 | $(\bar{1}2\bar{2})$ ± 0.84° |
| | 8 | 0.53, 462 | 8.28 ± 0.03 | $(112)$ ± 1.44° |
| | 9 | 0.57, 485 | 2.44 ± 0.24 | $(\bar{1}2\bar{2})$ ± 2.49° |
| | 10 | 0.675, 536 | 2.34 ± 0.01 | - |
| | 11 | 0.75, 567 | 2.10 ± 0.12 | $(11\bar{1})$ ± 1.65° |

In segment 1, the fatigue pre-crack opened fully at 149 N, followed by a minimal growth of 163 ± 6 nm and a more substantial extension of 3.00 ± 0.09 µm along the $(1\bar{1}\bar{1})$ ± 3.05° cleavage plane at applied 124 N and 0.09 mm. In this segment, DIC-based SIFs calculation showed a dominant mode I, with minimal mode II contribution, and the *J*-integral showed a steady linear increase until the crack opened and then proceeded with a steeper linear increase once the crack started growing into segments 2 and 3 (**Fig. 5**b).

In segment 2, the crack extended by 4.61 ± 0.30 µm under 183 N at 0.15 mm crosshead displacement, with propagation following a path between high local misorientations. This trend continued in segments 3 to 4, with extensions of 8.10 ± 0.01 µm and 4.59 ± 0.19 µm at 258 N and 312 N, respectively. In segments 2 to 4, the crack growth was influenced by local misorientation fields, which are characteristic of AA-5052, a non-heat-treatable aluminium alloy strengthened by cold work. As established from EBSD, this cold work results in a pronounced crystallographic texture and a structure with multiple subgrains having boundary misorientations ranging between 1.5° and 4.5°. High-resolution EBSD mapping of multiple regions across the specimen (**Fig. 3**b and c) reveals regions of intense type III plastic deformation, marked by low-angle grain boundaries, dislocation walls, and relatively high geometrically necessary dislocation density. Additionally, the X-ray measured residual type I elastic stress in the material is low, estimated at –20.50 ± 11.51 MPa, indicating that most of the deformation captured in the microstructure is plastic in nature. Notably, the linear increase in DIC-based mode I and the *J*-integral plateaued in segment 4, when the crack started to curve up to segment 5.

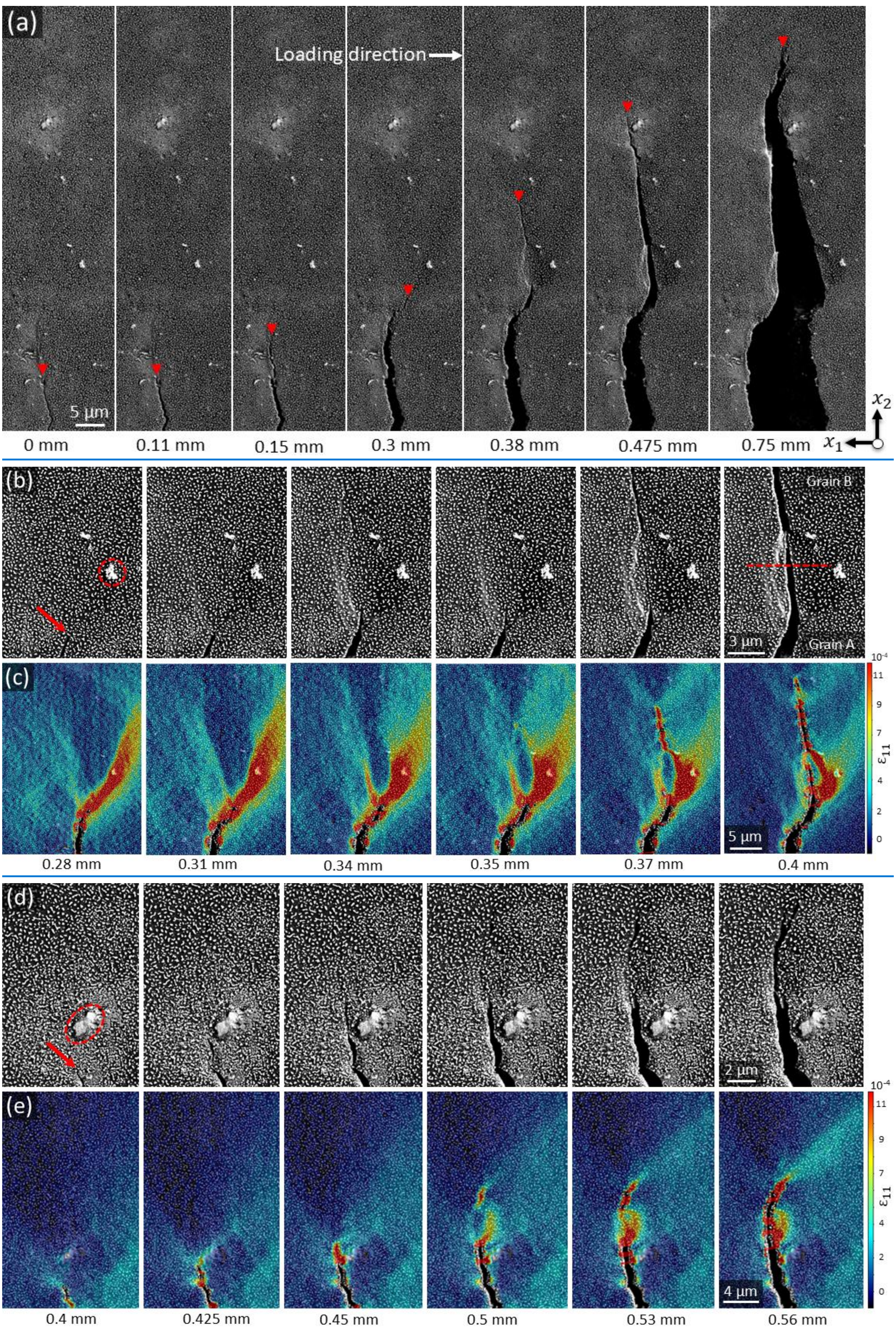

**Fig. 4. SEM and DIC characterisation of the crack growth.** (a) SEM image sequence showing the full progression of the crack from the initial fatigue pre-crack (0 mm) to its final length at 0.75 mm crosshead displacement. The red arrowheads track the advancing crack tip. (b) SEM image and (c) in-plane axial strain ($\varepsilon_{11}$) sequence showing crack propagation from Grain A into Grain B across a Σ13 grain boundary. The red arrow indicates the advancing crack tip, while the dashed circle highlights the Fe-rich precipitate. (d) SEM image and (e) in-plane axial strain ($\varepsilon_{11}$) sequence showing the crack tip interaction with a Fe-rich precipitate (see Fig. S4 for chemical analysis) during the transition from segment 7 to segment 8. For the strain field during crack propagation, see Movie S1.

In segment 5, the crack followed the $(\bar{1}\bar{2}1)$ ± 3.30° plane for 5.34 ± 0.02 µm, reaching an applied 335 N at 0.35 mm. At this point, the crack crossed into Grain B through a Σ13 ± 2.76° grain boundary. Analysis of the strain field and SEM imaging (**Fig. 4**b and c) revealed that, while the crack tip continued to open steadily, it encountered localised resistance at the boundary, causing a visible deflection in its path. The $\varepsilon_{11}$ strain field shows a marked concentration just before boundary crossing (0.28 – 0.31 mm crosshead displacement), which is redistributed upon entry into Grain B (0.34 – 0.37 mm crosshead displacement), consistent with local plastic accommodation during intergranular crack transfer. This transfer is also reflected in the DIC-based SIFs (**Fig. 5**b): mode II exhibited an increase approaching the boundary, while mode I remained dominant. The fluctuation in mode II suggests shear-driven deformation transfer, while the overall increase in the *J*-integral indicates an elevated energy requirement for intergranular propagation. Despite this, the crack advanced successfully into Grain B, as low-energy Σ13 boundaries pose limited resistance for damage transfer.

In segment 6, the crack showed a marked increase in crack length to 18.83 ± 0.28 µm along the (112) ± 1.44° plane at applied 372 N and 0.40 mm. In segment 7, the crack path transitioned to the $(\bar{1}2\bar{2})$ ± 0.84° plane, extending 2.34 ± 0.16 µm at applied 408 N and 0.45 mm. In both segments, there was a gradual linear increase in DIC-based mode I and *J*-integral, while mode II plateaued, indicating forward propagation, primarily under mode I conditions.

As the crack approached a Fe-rich intermetallic precipitate (see Fig. S4 for chemical analysis), associated with common Al–Fe–Mn–Si compounds in AA-5xxx series alloys, it began to deviate from its prior trajectory. The SEM and strain field sequence (**Fig. 4**d and e) clearly illustrates how the crack tip manoeuvred around the precipitate between 0.5 mm and 0.56 mm crosshead displacement, accompanied by a redistribution of $\varepsilon_{11}$ strain along the periphery of the particle and the strain localises along the new crack path, suggesting a surface-sensitive deviation driven by particle interaction and local plastic accommodation. During this segment, both DIC-based mode I and II SIFs (**Fig. 5**b) sharply decreased at 0.5 mm crosshead displacement, indicating a momentary reduction in crack driving force due to obstacle interaction.

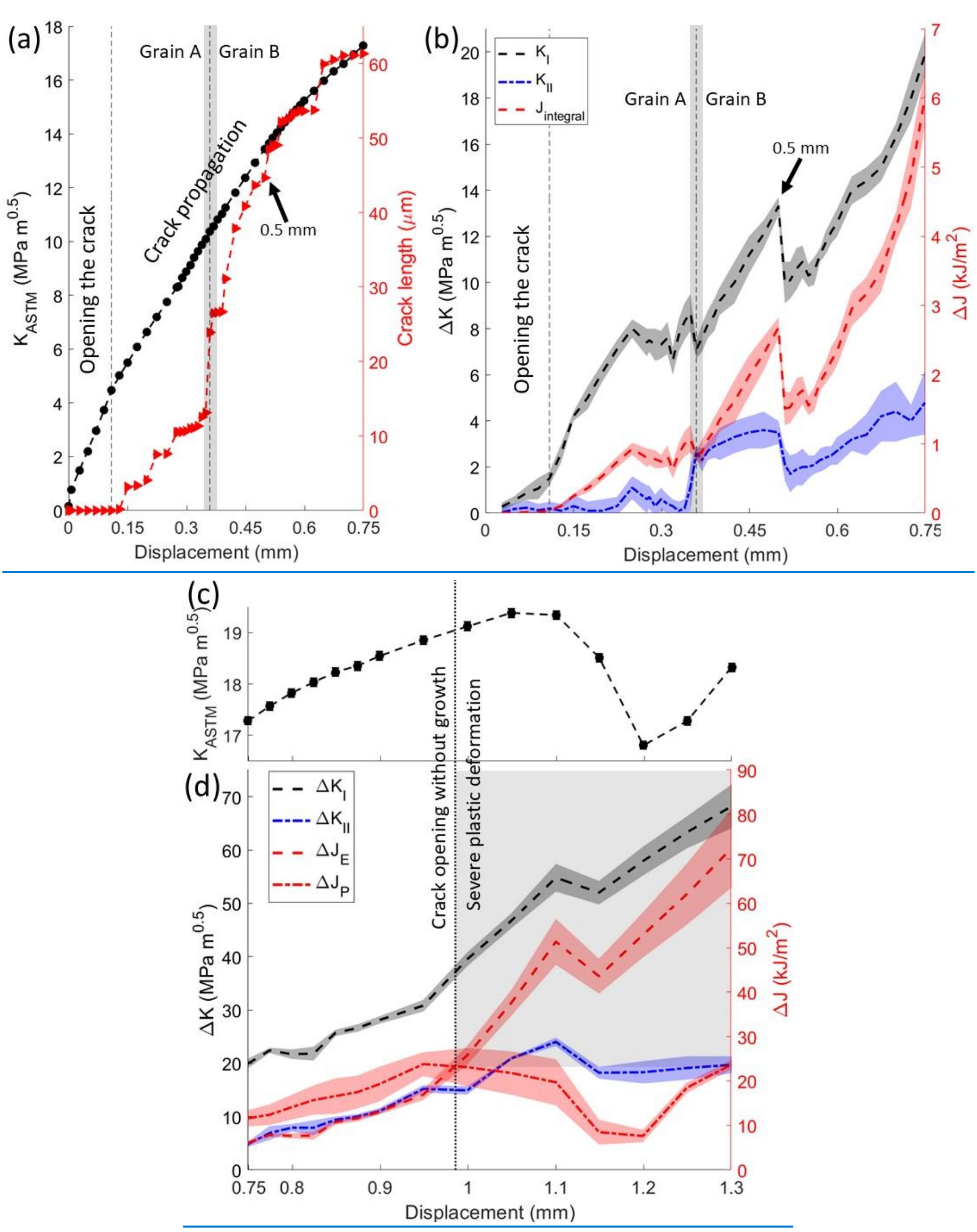

**Fig. 5. Crack stress intensity factors and strain energy release rate.** (a) ASTM-based mode I SIF ($K_{ASTM}$, black) plotted against displacement, alongside measured crack growth (red, right axis) up to when the crack stopped growing, as defined from **Fig. 4**a. (b) DIC-based fracture metrics, including mode I

(black), mode II (blue), and strain energy release rate ($\Delta J$, red, right axis) plotted against crosshead displacement for initial crack opening and stable crack propagation regimes. (c) $K_{ASTM}$ and (d) DIC-derived fracture metrics, including both elastic ($\Delta J_E$) and elastoplastic ($\Delta J_P$) components of the strain energy release rate (red, right axis), for crack opening without growth and severe plastic deformation regimes, as defined in **Fig. 1**a. Shaded bands around plotted curves represent domain-wise variability in fracture metrics ($\Delta J$, $\Delta K_I$, and $\Delta K_{II}$) calculated using the EDI method. These are derived from the converged integration region (see pink shaded area in **Fig. 2**d) and reflect standard visualisation of convergence spread in DIC-based analysis. The grey shaded region in D marks the regime beyond 0.984 ± 0.011 mm crosshead displacement, where the elastic assumptions of linear fracture mechanics break down. See Fig. S5 for the combined figures of A and C, as well as B and D.

In segment 8, the crack realigned with the (112) ± 1.44° plane and extended 8.28 ± 0.03 µm at nominal 462 N and 0.53 mm. The strain fields show a surface-driven crack path, influenced by stress intensification from the precipitate. The *J*-integral increased steadily throughout this segment, while DIC-based mode I and II, and the *J*-integral rose again. This behaviour suggests a recovery of crack propagation capacity after bypassing the precipitate, indicating resumed forward propagation under mixed-mode conditions.

In segment 9, the crack continued along $(\bar{1}2\bar{2})$ ± 2.49° for 2.44 ± 0.24 µm at nominal 485 N and 0.57 mm, and in segment 10, the crack exhibited non-crystallographic propagation over 2.43 ± 0.01 µm at nominal 536 N and 0.675 mm. Finally, in segment 11, the crack followed the $(11\bar{1})$ ± 1.65° plane for 2.1 ± 0.12 µm at nominal 567 N and 0.75 mm, where the crack terminated within Grain B after 63.65 ± 1.23 µm growth under displacement control (**Fig. 4**a). During segments 9 to 11, mode I and II, and the *J*-integral continue to increase, with mode I being dominant.

Overall, given the notch size, the ASTM-based SIF estimate (**Fig. 5**a and c) remained proportional to the applied load and insensitive to crack–microstructure interactions. In contrast, analysis of the in-plane displacement fields obtained from DIC enabled calculation of the local strain energy release rate (*J*-integral) and mode-separated SIFs along the crack path, with the *J*-integral and the decoupled mode I and II exhibiting strong sensitivity to local microstructural conditions and corresponding well with the observed crack growth.

### ~~2.2.~~3.2. Ductile tearing

Beyond 0.75 mm crosshead displacement, further displacement resulted in a crack opening but with no further propagation. From SEM images (**Fig. 6**a), two distinct regimes were identified: one where the crack opened with no apparent plasticity, and another where pronounced plastic deformation became apparent, evidenced by the formation of slip bands emitted from the crack tip. While initial visual inspection suggested this transition occurred at a crosshead displacement of 1 mm; quantitative analysis identified the transition as the first loading increment at which (i) crack length remained constant under increasing displacement and (ii) the divergence between the $J$-integrals – calculated using elastic properties to extract mode I and II (i.e., $\Delta J_E$) and the $J$-integral calculated using elastoplastic properties (i.e., $\Delta J_p$) – revealed a more accurate transition point at 0.984 ± 0.011 mm crosshead displacement and 23.28 ± 2.81 kJ/m$^2$ (**Fig. 5**d).

Note that the small differences visible between $\Delta J_E$ and $\Delta J_P$ at ~0.75 mm displacement reflect methodological and numerical factors rather than actual plasticity, because inverse crystal plasticity solutions are inherently more sensitive to measurement noise and convergence behaviour than purely elastic reconstructions, which affects $J$-integral convergence (see the larger variability in **Fig. 5**d compared to the elastic case) and leads to minor deviations in calculated stress and energy fields even for elastically responding cracks (Supplementary Information, Section 1). Thus, the divergence between $\Delta J_E$ and $\Delta J_P$ becomes physically meaningful only after crack arrest, when plastic deformation develops, as shown in Fig. 5d and discussed in this section. Beyond ~0.7 mm and up to the convergence region, $\Delta J_E$ and $\Delta J_P$ exhibit similar trends, although $\Delta J_P$ aligns more closely with $K_{ASTM}$ (based on applied load). This reinforces that $\Delta J_P$ only gains physical significance after crack arrest, coinciding with the onset of plasticity.

The energy-based transition confirms the visual observations of the onset of severe plastic deformation at the crack tip and marks the breakdown of the elastic assumptions required for valid mode I and II calculations. However, beyond 0.984 mm crosshead displacement, using elastic properties beyond this point has shown – qualitatively – that mode I remains dominant and is increasing, while mode II plateaued, signalling a regime shift from mixed-mode microstructure-sensitive crack to mode I (load)-dominated crack. In contrast, the elastoplastic $J$-integral ($\Delta J_p$) began to plateau with applied displacement, reflecting the onset of severe plastic deformation, and matching the nominal force behaviour, including the force drop around 1.2 mm (**Fig. 1**e). As the test was conducted under displacement control, this drop in load is due to a reduction in structural stiffness due to the onset of plastic deformation near the arrested crack tip. At this stage, the crack ceases to grow, and the surrounding material begins to yield, leading to a localised structure softening.

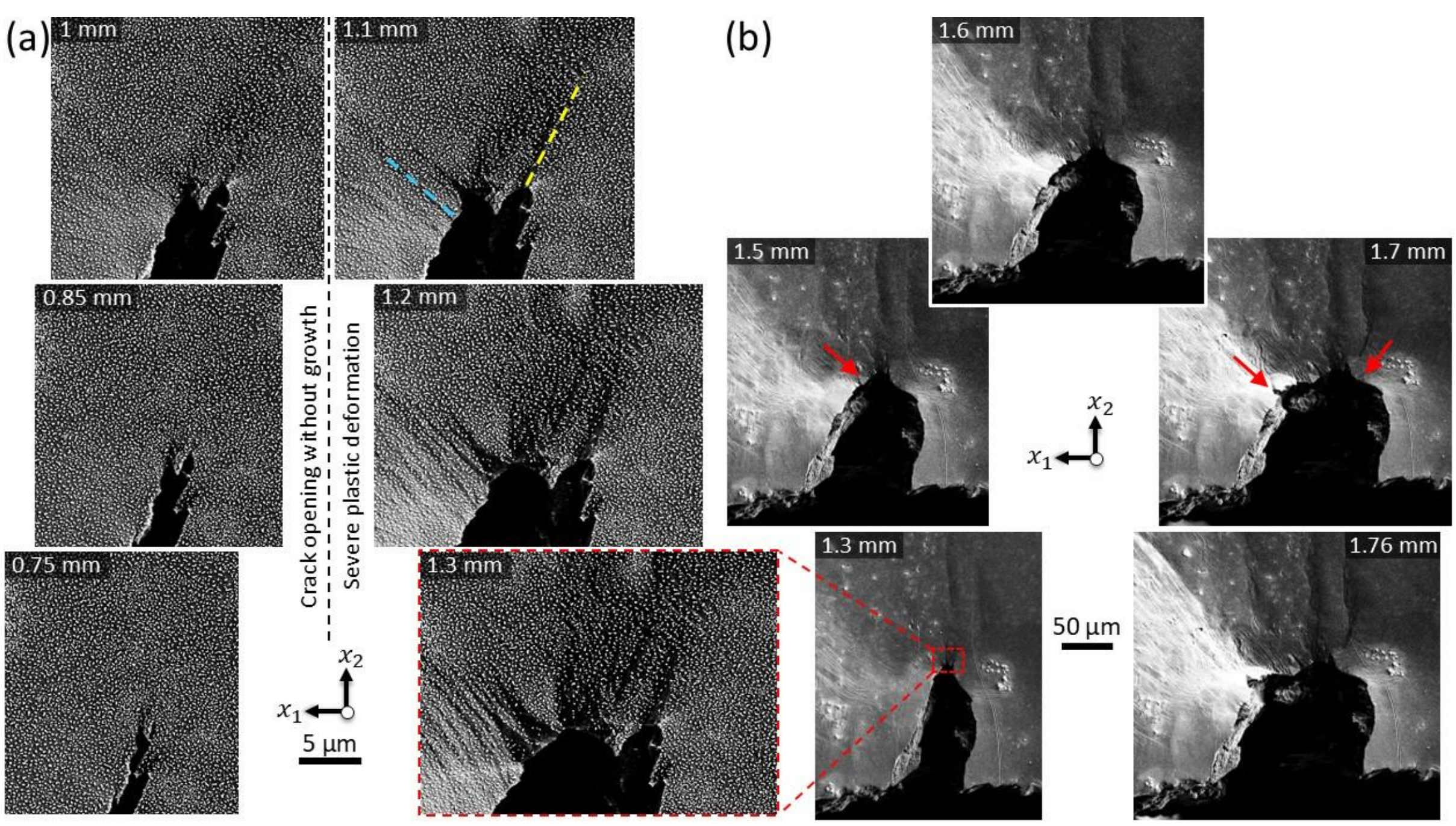

**Fig. 6. Plasticity development at the crack tip.** (a) SEM sequence showing crack behaviour from final propagation (0.75 mm) through to severe plastic deformation (1.3 mm). The yellow and blue dashed lines are $(1\bar{1}\bar{1})[0\bar{1}1]$ and $(\bar{1}1\bar{1})[0\bar{1}\bar{1}]$ slip trace, respectively (b) Wider field SEM views show continued deformation from 1.3 mm to 1.8 mm crosshead displacement. Red arrows indicate ductile tearing and localised surface shear.

As displacement increases, initially $(1\bar{1}\bar{1})[0\bar{1}1]$ (yellow dashed line) and $(\bar{1}1\bar{1})[0\bar{1}\bar{1}]$ (blue dashed line in **Fig. 6**a) slip bands radiated from the crack tip and blunting initiates, forming a plastic stretch zone. After 1.3 mm of crosshead displacement, the crack was unloaded as shown in **Fig. 1**e, with the crack remaining open at 0.766 mm crosshead displacement. The crack was then loaded again (**Fig. 7**a), and despite continued crosshead displacement from 1.3 mm to 1.76 mm, the load plateaued around 640 N, indicating crack arrest under severe plastic strain conditions, absorbing the applied energy rather than contributing to further elastic stress buildup. In addition, as crosshead displacement increased, slip band formation and ductile tearing emerged, particularly beyond 1.5 mm (highlighted by red arrows in **Fig. 6**b). The crack tip blunted into a rounded geometry, and the surrounding surface developed an elliptical singularity, suggesting the development of a stretch zone and plasticity-dominated blunting. These features align with macro-scale crack opening, indicating irreversible energy dissipation via plastic flow, and confirm the shift to opening-mode-controlled deformation in a macroscale plastic regime. The crack remained open upon unloading, confirming substantial residual strain and irreversible deformation in the surrounding matrix.

Thus, the divergence between $\Delta J_E$ and $\Delta J_P$ – or linear elastic fracture mechanics (LFEM) to elastic-plastic fracture mechanics (EPFM) – marks a shift from microstructure-sensitive propagation to macro-scale plastic opening, and the formation of a plasticity-driven process zone, where the material yields locally and redistributes stress, and further displacement is accommodated not by increased stress or crack advance, but by plastic flow and blunting at the crack tip. This is a hallmark of ductile fracture, and its onset aligns with the plateau in mode II contribution as the crack aligns with the principal stress direction, with crack arrest corresponding to full ligament yielding before crack advance as a long crack, where global loading and plastic dissipation – not microstructural barriers – govern the fracture response.

Further evidence of plasticity-driven fracture behaviour after crack arrest is revealed by the topography and high-magnification SEM imaging in **Fig. 7**c showing the blunted crack tip and extensive plastic surface features, including slip steps and shear bands. The 3D surface height map (**Fig. 7**b) shows an elliptical depression centred at the crack tip and approximately 180 µm deep at its centre, extending ahead of the arrested crack tip, and indicates continued localised plastic deformation and stretch-zone formation under plane stress. This residual deformation confirms that, after crack arrest, applied energy is dissipated through local plastic flow rather than further crack advance, consistent with the SEM-DIC strain-field evolution.

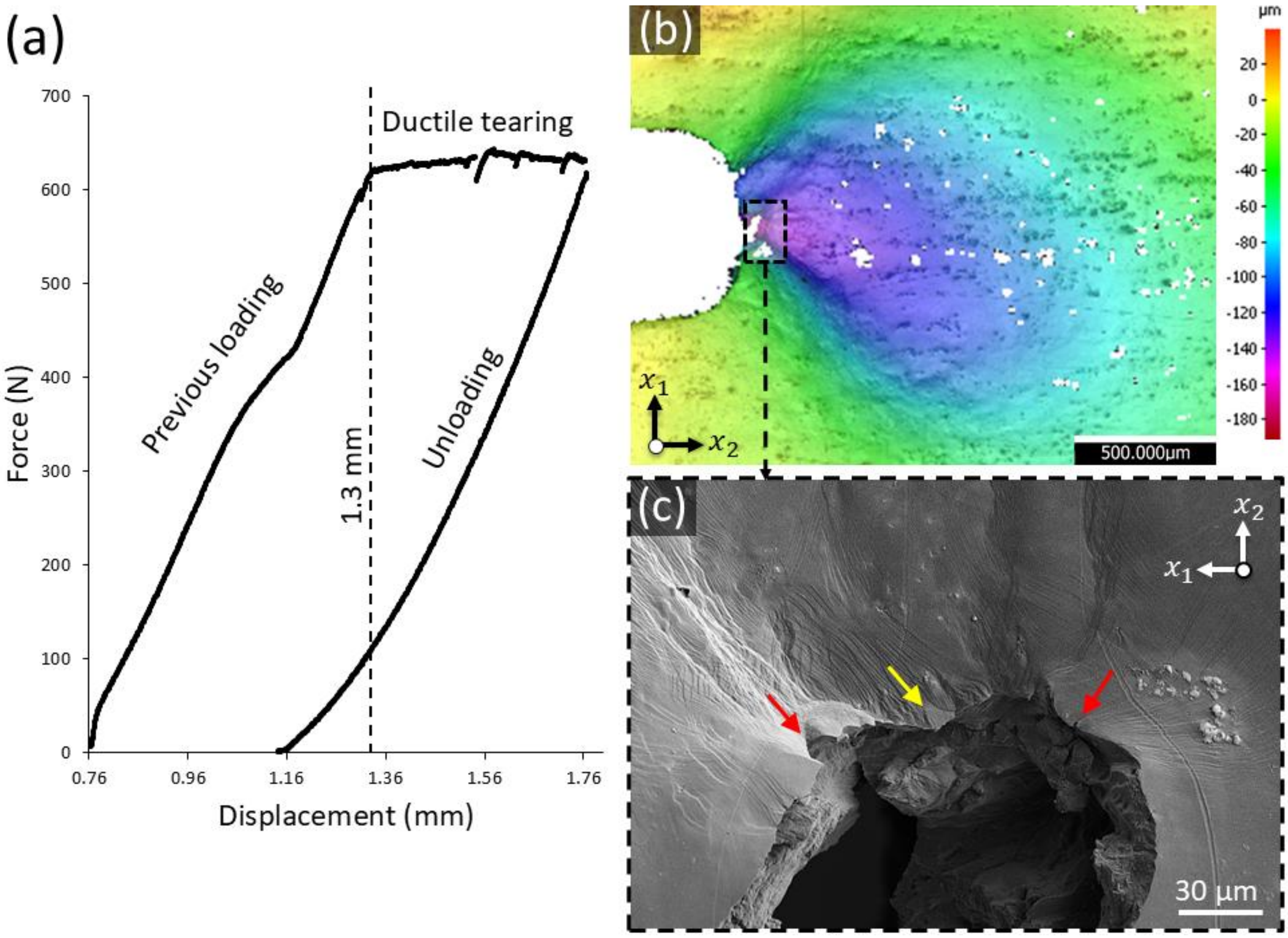

**Fig. 7. Localised plasticity at the crack tip.** (a) The nominal load-displacement curve was recorded during reloading after crack arrest. (b) 3D topography map of the post-arrest crack region. (c) High-resolution SEM image of the same region. Red arrows denote sites of ductile tearing where secondary cracks nucleated, while the yellow arrow highlights a region where intersecting slip bands and localised contrast features suggest the possible formation of a sessile Lomer dislocation lock. Arrows are spatially correlated with the features identified in **Fig. 6**b.

These observations are interpreted quantitatively through the evolution of local fracture metrics and process-zone development, which form the basis for the mechanistic discussion that follows.

## ~~3.~~4. Discussion

The central physical result of this study is not the observation that ductile cracks eventually arrest due to plasticity, which is well established, but the direct experimental quantification of how the crack-tip driving force evolves during this transition. By reconstructing both elastic and elastoplastic energy measures from the same measured displacement fields, we demonstrate that crack arrest coincides with a measurable change in energy partitioning: from predominantly recoverable elastic energy associated with crack advance to increasing plastic dissipation associated with crack blunting. This transition is detected independently of crack length and correlates directly with observed slip-band emission and process-zone expansion. The contribution of the present work is therefore the experimental linkage between crack-tip field evolution and the onset of plasticity-dominated arrest under controlled microstructural conditions.

Below we will interpret the experimental observations in terms of microstructure-sensitive propagation, and crack-tip driving force evolution and process-zone development.

### ~~3.1.~~4.1. Microstructure-sensitive crack propagation

AA-5052 is a non-heat-treatable alloy that relies on solid solution strengthening and strain hardening rather than precipitation hardening [39,40]. In this alloy with pronounced microstructural plasticity, we observed that the crack propagated with limited to no observable plasticity at the crack tip, in quasi-brittle[*] matter along non-cleavage planes, with minimal observable plasticity at the tip. Strain fields remained highly localised at the crack tip but confined to a small process zone ahead of the crack. Additionally, our results reveal that observed crack propagation is highly influenced by the surface microstructural heterogeneities, local plastic deformation, and the principal stress direction rather than crystallographic constraints alone, as the crack propagated along mostly non-cleavage planes. Although dynamic GND evolution at the crack tip could not be measured, the SEM-DIC strain fields reveal that early crack propagation occurs within a process zone comparable to or smaller than this heterogeneity scale. Mixed-mode fluctuations observed in **Fig. 5**b are therefore interpreted as arising from local compatibility stresses and slip heterogeneity associated with the initial subgrain structure.

While mode I – the principal loading mode – remains the dominant fracture mechanism throughout the microstructure-sensitive crack growth, mode II contributions become non-negligible in segments where the crack interacts with grain boundaries and precipitates, with possible mode III contributions that cannot be resolved experimentally.

[*] Quasi-brittle crack propagation occurs when fracture is influenced by microstructural heterogeneities, leading to a finite process zone with distributed damage ahead of the crack tip. Unlike purely brittle fracture, it exhibits gradual crack growth, size-dependent strength, and sensitivity to local material disorder [98,99]. Here, the absence of observable slip bands, crack tip blunting, or plastic wake during crack propagation, combined with the highly localized strain field, supports the classification of this regime as quasi-brittle.

Comparatively, ASTM-based SIF estimates scaled with load and ignored microstructural effects, while DIC-derived *J*-integrals and mode-separated SIFs were highly sensitive to local microstructure and aligned with crack growth behaviour. Nonetheless, during the microstructure-sensitive crack growth, the ASTM and DIC-based SIFs analysis reveals that the crack propagated at a stress intensity factor substantially lower than the typical fracture toughness range reported for AA-5052 (25–30 MPa·$m^{0.5}$ [74]).

### ~~3.2.~~4.2. Microstructure-sensitive to load-dominated crack transition

As the crack stopped growing, the transition from quasi-brittle to plasticity-dominated crack behaviour was governed primarily by surface-local phenomena, as evidenced by SEM-DIC strain fields and the development of a crack-tip plastic zone from initial blunting and surface slip-band formation (**Fig. 6**a) to the evolving surface topography (**Fig. 7**b), such that additional applied energy is dissipated through plastic flow rather than crack advance, prior to the onset of plasticity-dominated blunting and ductile tearing (**Fig. 7**c). These observations are typical characteristics of crack arrest in ductile alloys under monotonic loading [19,22,75].

Traditionally, microstructure-sensitive crack propagation in ductile alloys at low energy has often been discussed in terms of crack length effects [62,76,77] (i.e., short crack), whereas a substantial theoretical body of work also emphasises the controlling role of crack-tip process-zone development and plastic shielding [78,79]. In the present experiments, when the process zone is small, the crack interacts with only one grain, promoting quasi-brittle, crystallographic facet growth even in otherwise ductile materials (e.g., cleavage-like facets in AA-5052 [75]). As the *J*-integral increases, the process zone expands to engage multiple grains and activates mechanisms such as slip-band interaction, crack deflection, and microvoid coalescence. Consistent with process-zone-based interpretations of microstructure-sensitive crack behaviour, the transition observed here corresponds to the process zone reaching a critical extent rather than to a unique crack length.

In this view, the onset of plastic dissipation is indicated by a clear divergence between the elastic and elastoplastic energy measures extracted from the displacement fields – i.e., $\Delta J_{crit}$ when $\Delta J_E \geq \Delta J_P$ – coincident with the cessation of crack advance (Fig. 4a). It is worth emphasising that the observed condition $\Delta J_E \geq \Delta J_P$ (**Fig. 5**d) after the onset of ductile tearing is a natural consequence of the analysis framework employed. In both cases, the same experimentally measured SEM-DIC displacement field is imposed as a boundary condition. The purely elastic formulation cannot accommodate stress relaxation and therefore yields a stiffer response, leading to an overestimation of the recoverable strain energy and hence $\Delta J_E$ (demonstrated further in Fig. S6b). In contrast, the elastoplastic formulation allows local stress relaxation through plastic flow, causing the incremental elastoplastic energy release rate $\Delta J_P$ to saturate despite continued deformation. Because the present analysis is based on incremental *J*-integral values, this divergence reflects the transition from energy being available for crack advance to energy being dissipated through plastic deformation. The condition $\Delta J_E \geq \Delta J_P$ therefore provides a physically consistent indicator of plasticity-dominated crack arrest rather than a contradiction of fracture mechanics expectations.

Furthermore, the mechanistic basis of $\Delta J_{crit}$ is supported by direct physical observations: visible slip bands initiate ahead of the tip (**Fig. 6**a), the crack blunts (**Fig. 6**b), and a growing plastic zone develops (**Fig. 7**b and c). Accordingly, in the present experiments, the observed $\Delta J_{crit}$ marks the displacement at which elastic and elastoplastic energy measures diverge and plastic dissipation dominates. Its numerical value is conditional on material, microstructure, and constraint, but its identification demonstrates the existence of an experimentally measurable energy-partition transition governing crack arrest under the present conditions.

The present observations can be interpreted in light of process-zone-based perspectives, such as the Kysar framework [79], which emphasises the competition between energy dissipation by new surface formation and by dislocation-mediated blunting at the crack tip. Here, the spatiotemporal evolution of the process zone (Movie S1) shows strain fields progressively radiating from the crack tip under increasing displacement, highlighting gradual enlargement of the plastically deforming region, particularly as the crack arrests. The SEM-DIC-based $\Delta K_I$, $\Delta K_{II}$, and $\Delta J$ therefore provide physically grounded measures of crack-tip driving force that act as practical scalars for process-zone development at the surface.

Mathematically, for material under plane stress and light to moderate strain hardening, the process zone size ($r_p$, visualised in Fig. 8) can be estimated from the *J*-integral using [80]:

$$r_p \approx \frac{2}{\pi}\frac{J}{\sigma_0} \quad (13)$$

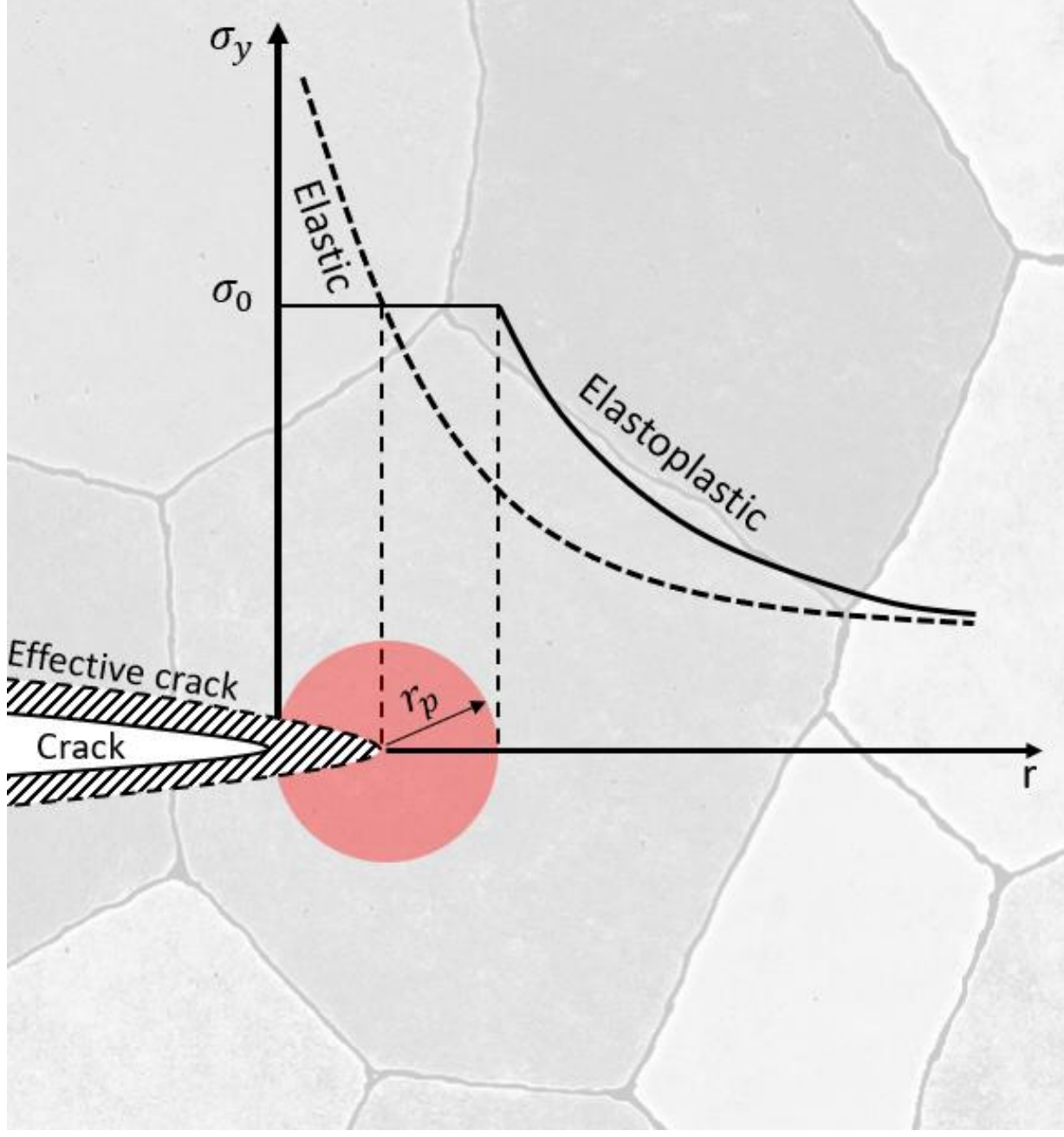

Fig. 8: Visualisation of the process zone and effective crack length at the grain-scale level for a microstructure-sensitive crack.

In our study, uniaxial tensile testing of a similarly cold-worked and textured AA-5052 specimen yielded a 0.2% offset yield stress ($\sigma_0$) of 193 MPa (see Supplementary Fig. S6a). Using this value in equation (13), the critical energy threshold identified in our measurements, $\Delta J_{crit}$ = 23.28 ± 2.81 kJ/m$^2$, corresponds to a process zone size on the order of 76.8 ± 9.3 µm at the point of transition. Notably, this process zone spans more than twice the average grain size of the material, which was measured at 31.7 ± 8.0 µm. This dimensional comparison supports the interpretation that, once the plastic zone spans multiple microstructural heterogeneities, the influence of individual GND-rich regions diminishes, and crack behaviour becomes increasingly load-dominated. Notably, this dimension closely matches the radial extent of the unindexed region observed in serial sectioning 3D EBSD of the arrested crack (Fig. S8e), where severe lattice distortion prevented reliable orientation reconstruction. The spatial agreement between the energetically estimated process-zone size and the diffraction-degraded volume provides independent structural support for the interpretation that crack arrest coincides with intense plastic strain localisation spanning multiple grains. While evolving GND accumulation at the crack tip was not directly measured, the observed slip-band emission and large strain gradients (**Fig. 6**) confirm that plastic strain localisation surpasses the initial stored deformation state at this stage.

Additionally, although the initial pre-crack was relatively large, 76.42 ± 0.70 µm, it exhibited microstructure-sensitive crack behaviour following stress-relief annealing, indicating that crack length alone did not govern the transition. Instead, the onset of long-crack behaviour coincided with the development of a process zone that expanded to engage multiple grains, and only then did the fracture behaviour become less sensitive to grain-scale heterogeneity, marking the transition toward a nominally load-dominated regime.

This indicates that, under the present free-surface and plane stress conditions, the observed transition is not correlated with crack length alone but coincides with the expansion of the surface-resolved plastic process zone relative to the microstructure. In such cases, the process zone is the controlling length scale for fracture behaviour.

Similar process-zone-controlled transitions have been reported across various alloy systems, including aluminium [22,75,81], titanium [34], Ni-based superalloys [82], and steels [83,84], where crack behaviour changes once the crack-tip process zone expands beyond microstructural length scales.

Taken together, these findings are consistent with the interpretation that the transition from microstructure-sensitive crack behaviour to a plasticity-dominated and nominally load-controlled regime is process-zone governed. Mixed-mode quasi-brittle fracture arises when the process zone is small, whereas plasticity-dominated behaviour emerges once it expands sufficiently to engage multiple grains. This framework suggests the value of complementing length-based metrics (e.g., Δa) with plasticity-driven descriptors (e.g., $r_p(t)$) when interpreting crack resistance in ductile alloys. Further studies across FCC systems with varying hardening responses and microstructures will be required to evaluate the broader generality of the $\Delta J_{crit}$-based transition framework.

Nonetheless, it is worth noting that the present process-zone estimates are derived from surface-resolved displacement fields and therefore reflect the free-surface response under plane stress conditions. Subsurface plasticity and through-thickness constraint may evolve differently and could contribute to the overall crack-front driving force. While the surface transition identified here coincides with crack arrest under the tested geometry and thickness, full three-dimensional characterisation would be required to determine how this transition scales with thickness and constraint. The present conclusions, therefore, pertain to the experimentally accessible surface process zone rather than to a fully three-dimensional crack-front field.

### ~~3.3.~~4.3. Cautionary note on *J*-integral applicability

It is important to distinguish between Δ*J* values extracted during active crack growth and those obtained after crack arrest. During crack propagation, crack-tip deformation remains localised and surface dominated, justifying the use of surface-resolved fracture metrics to characterise the transition. After crack arrest, extensive plastic deformation develops and through-thickness constraint effects become significant. Δ*J* values extracted in this post-arrest regime are therefore surface specific and are used here only to indicate the breakdown of elastic assumptions, the transition from mixed-mode microstructure-sensitive crack to mode I (load) governed crack, and the onset of plasticity-dominated blunting (**Fig. 5**d), rather than to represent crack-front driving forces. Additionally, although the *J*-integral is an appropriate metric for studying microstructure-sensitive cracks, its validity in elastoplastic materials requires that the integration domain fully encloses the plastic zone [69,85,86]. Its successful application in this study relies on the near-homogeneous elastic anisotropy of the grains, with stiffness differences below 3%, which minimised the influence of grain misorientation.

More generally, the physical meaning of the *J*-integral at the micro-scale remains weakly demonstrated [15]. The present results show that it can provide a meaningful measure of the crack-tip field when plasticity anisotropy is effectively homogeneous, consistent with prior work in anisotropic single-phase materials [87–89]. Caution is required when extending this approach to polycrystalline systems with strong elastic or plastic anisotropy or sharp interfaces, where path independence may break down and interaction- or interface-integral formulations will be required [90–94].

While such approaches are well developed numerically, in situ experimental validation at the micro-scale remains constrained, especially for three-dimensional measurements. Existing synchrotron-based techniques, such as Laue micro-diffraction [95], differential aperture X-ray microscopy [96], and dark-field X-ray microscopy [97], are limited to elastic strains and post-mortem analysis, and volumetric correlation methods lack the necessary spatial resolution. As a result, surface-based SEM-DIC remains the only practical experimental method for resolving crack-tip mechanics during microstructurally sensitive short crack propagation in aluminium alloys.

## 5. Conclusion

This work delivers a direct, high-resolution experimental evidence of how fracture in cold-worked AA-5052 transitions from microstructure-sensitive crack growth to plasticity-driven, load-controlled arrest. By integrating SEM-DIC and EBSD with crystal plasticity finite element modelling, we quantified local crack-tip fields and extracted elastic (mode I and II) and elastoplastic energy release rates ($\Delta J_E$ and $\Delta J_P$) from measured displacement data.

Our findings reveal that crack behaviour is not governed by crack length alone but by the evolution of the crack-tip process zone relative to the microstructure. When the process zone is confined within a single grain, crack growth is mixed-mode and highly sensitive to local heterogeneities. As loading increases and the process zone expands to span multiple grains, plastic deformation dominates, microstructural constraints diminish, and the crack aligns with the macroscopic stress direction. This transition is marked by a divergence between elastic and elastoplastic energy measures ($\Delta J_E \geq \Delta J_P$), coinciding with crack-tip blunting, slip-band formation, and ductile tearing. These results establish an energy-based criterion for crack arrest and underscore the importance of process-zone size – not crack length – as the controlling length scale for fracture in ductile alloys.

While the numerical value of the transition energy is specific to the present material and constraint conditions, the methodology enables direct quantification of this transition and offers a framework for systematically investigating its dependence on microstructure and loading. This insight provides a mechanistic foundation for damage-tolerant design strategies and highlights the value of coupling in situ experimental methods with crystal plasticity modelling for predictive fracture analysis.

**Acknowledgement:** We thank Ms Rachel Willia (National Physical Laboratory) for the XRD measurements, Duaa Salim (University of Khartoum) for validating crack length measurements, Dr Ken Mingard (National Physical Laboratory) for reviewing the manuscript and Professor James Marrow (University of Oxford) for the in-depth discussions, which enhanced the quality of the article.

**Funding:** This work was supported by the National Measurement System (NMS) programme of the UK Department for Science, Innovation and Technology (DSIT), and the Royal Society Short Industrial Fellowship (SIF\R2\242005).

**Author Contributions: AK**: Conceptualisation, methodology, investigation, formal analysis, visualization, and writing - original draft. **BS**: Software, validation, and writing - original draft. **CG**: Investigation and visualization. **FD**: Writing - review & editing, supervision, resources, and funding acquisition.

**Competing Interests:** The authors declare no competing interests.

**Data and Materials Availability:** All data needed to evaluate the conclusions in the paper are present in the paper and the Supplementary Materials. Raw and processed datasets generated during the current study are available from the corresponding author on reasonable request. The MATLAB-based toolbox used in this study is available under an MIT license at https://doi.org/10.5281/zenodo.6411605.

# Supplementary Materials for

## In situ elucidation of mechanisms governing crack transition to plasticity arrest

Abdalrhaman Koko [1,2]*, Bemin Sheen [2], Caitlin Green [1], and Fionn Dunne [2]

[1] National Physical Laboratory, Hampton Road, Teddington TW11 0LW, United Kingdom

[2] Department of Materials, Royal School of Mines, Imperial College London, UK

* Corresponding author. E-mail: abdo.koko@npl.co.uk

## 1. Crystal plasticity model validation

To validate the implementation of the crystal plasticity UMATs in a 3D plane stress model, a comparison has been drawn in Figure S1 against a 2D plane stress model, composed of CPS4 elements. In both sets of example models, a single-orientation hcp material with the elastic properties, defined in equation (1), is used due to its appreciably higher stiffness in the c-axis direction. The ordering is such that the entries of the leading diagonal follow the order: $C_{11}$, $C_{22}$, $C_{33}$, $C_{12}$, $C_{13}$, $C_{23}$. The 3D model is 3 nm thick with three elements across the thickness. In all orientations, the $\sigma_{22}$ fields display a strong agreement between the 2D and 3D models. It should be noted that minor differences are to be expected, as the 3D model requires extrapolation of stress from integration points onto the front surface.

$$C = \begin{bmatrix} 162 & 92 & 69 & 0 & 0 & 0 \\ & 162 & 69 & 0 & 0 & 0 \\ & & 180 & 0 & 0 & 0 \\ & & & 35 & 0 & 0 \\ & & Sym. & & 46 & 0 \\ & & & & & 46 \end{bmatrix} GPa \tag{1}$$

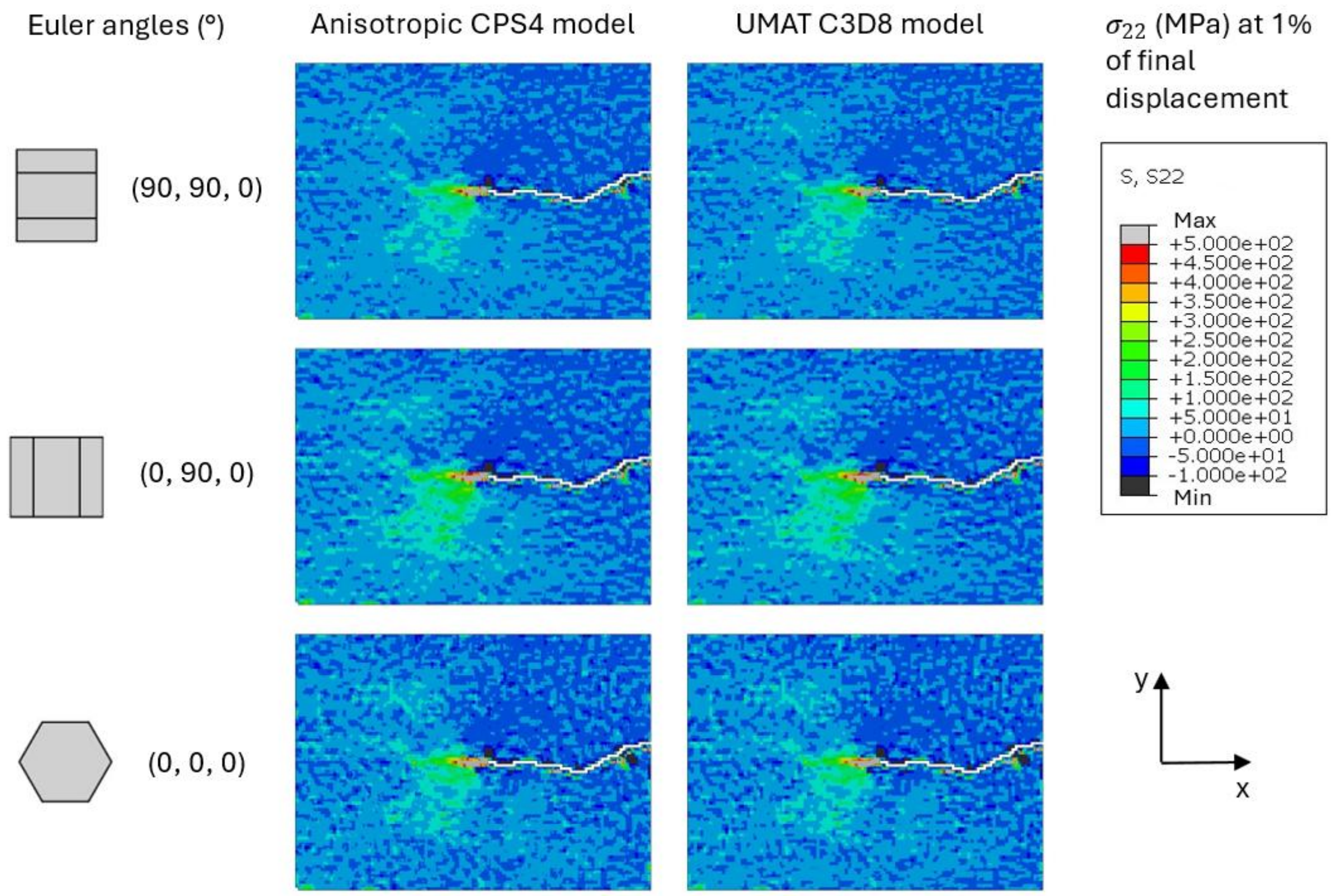

Figure S1: Example $\sigma_{22}$ stress fields produced by imposing displacements from DIC on a single-crystal model

The $\sigma_{22}$ fields of the (0°, 90°, 0°) orientation are further compared along the line profiles defined in Figure S2. Data from the line profiles A-A', B-B' and C-C' is presented in Figure S2 A to C, were we found an excellent correlation between the line profiles.

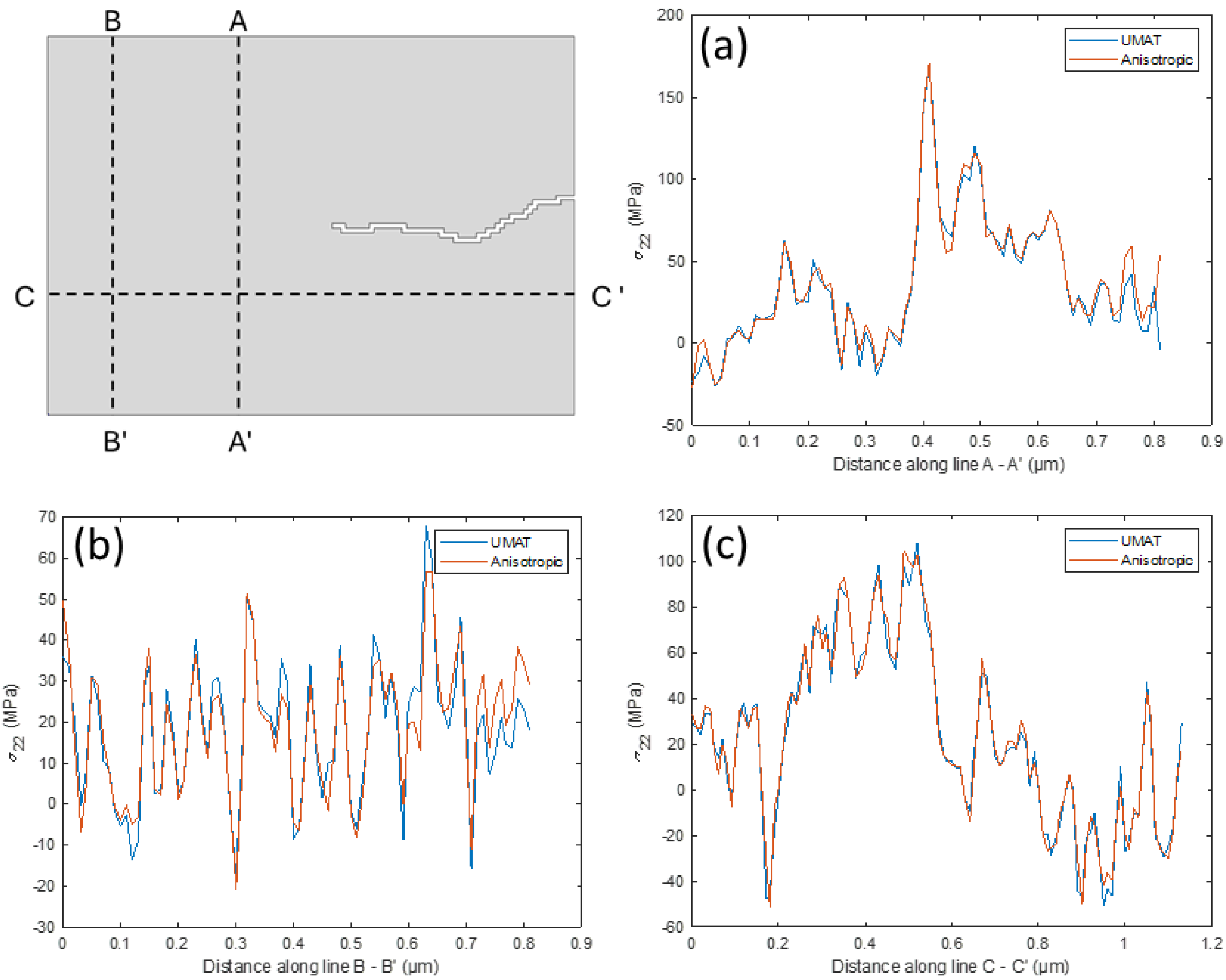

Figure S2: Locations of $\sigma_{22}$ line profiles

## 2. Supplementary figures

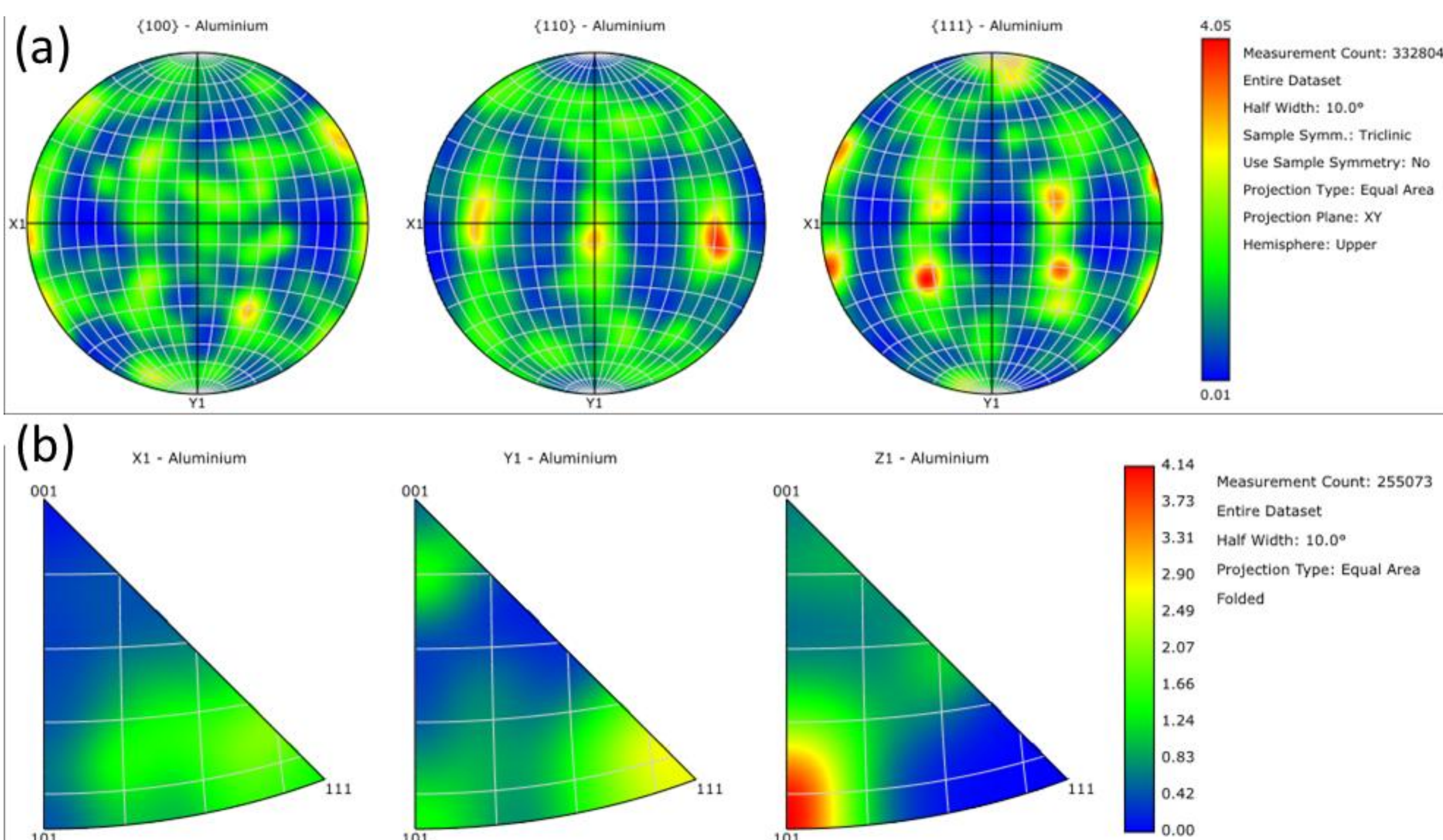

**Fig. S3. Specimen aggregate grain orientation.** (a) Pole figures for the {100}, {110}, and {111} planes of AA-5052, showing the crystallographic texture distribution with respect to the specimen's X1–Y1 reference frame. The data were collected using EBSD, and without applying specimen symmetry and plotted using equal area projections, highlighting the presence of deformation-induced preferred orientations along the {110} planes, consistent with cold-worked FCC aluminium alloys. (b) Corresponding inverse pole figures (IPFs) along the X, Y, and Z specimen directions, illustrate the orientation distribution of grains relative to the specimen axes. The IPF maps confirm a strong alignment along the {110}, indicative of the rolling texture commonly observed in strain-hardened aluminium alloys.

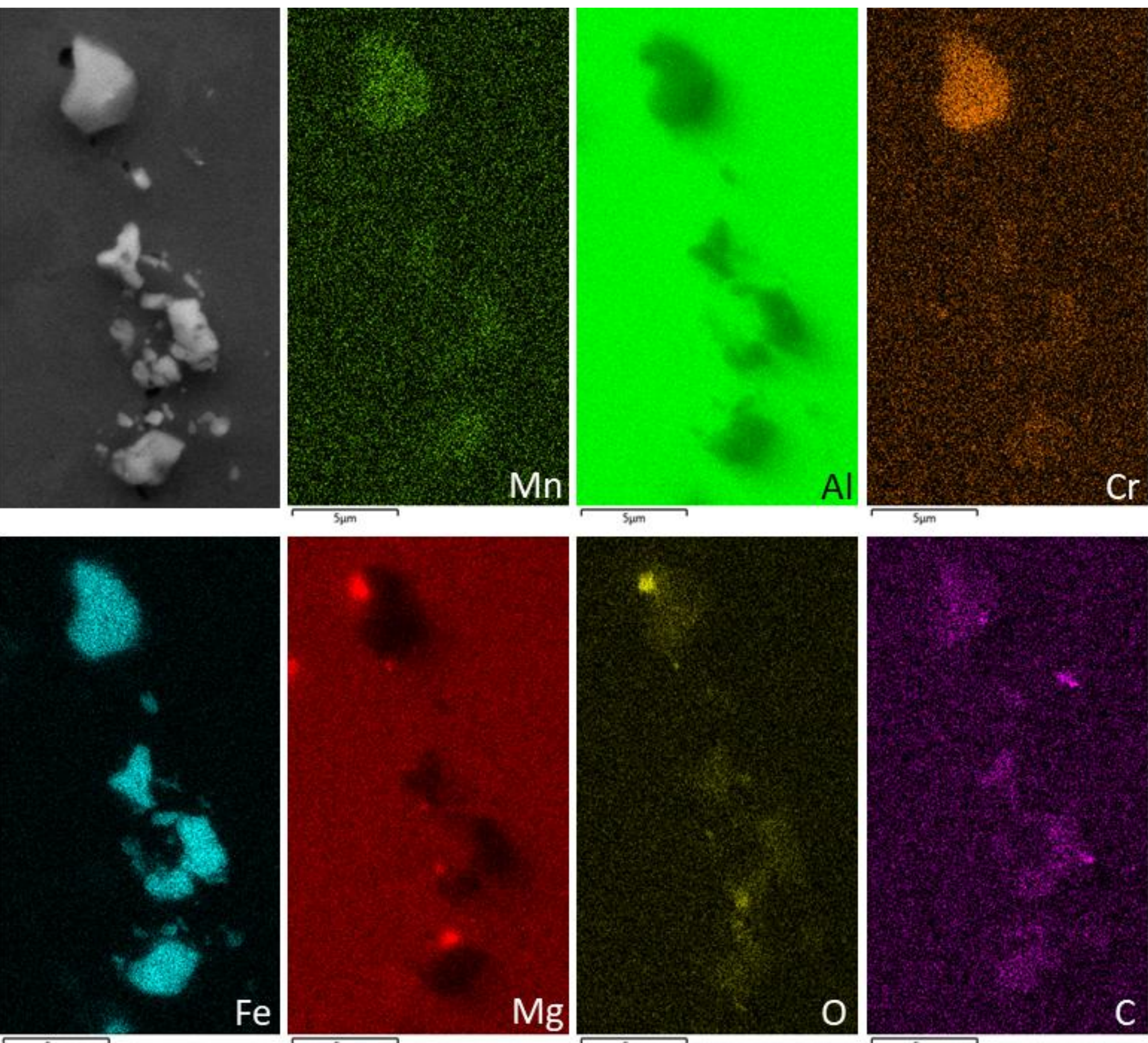

**Fig. S4. Precipitates chemical composition.** Backscattered electron (BSE) image (top left) and energy-dispersive X-ray spectroscopy (EDS) elemental maps of a Fe-rich intermetallic particle embedded in the AA-5052 matrix. Elemental distribution maps reveal strong enrichment in Fe and co-localised signals of Mn and Cr, suggesting the presence of an Al–Fe–Mn–Cr intermetallic phase. The precipitate also shows localised Mg and O, possibly related to surface oxidation or minor phase segregation. The Al map shows matrix depletion in the particle region, while trace carbon is likely due to surface contamination. Scale bars: 5 µm.

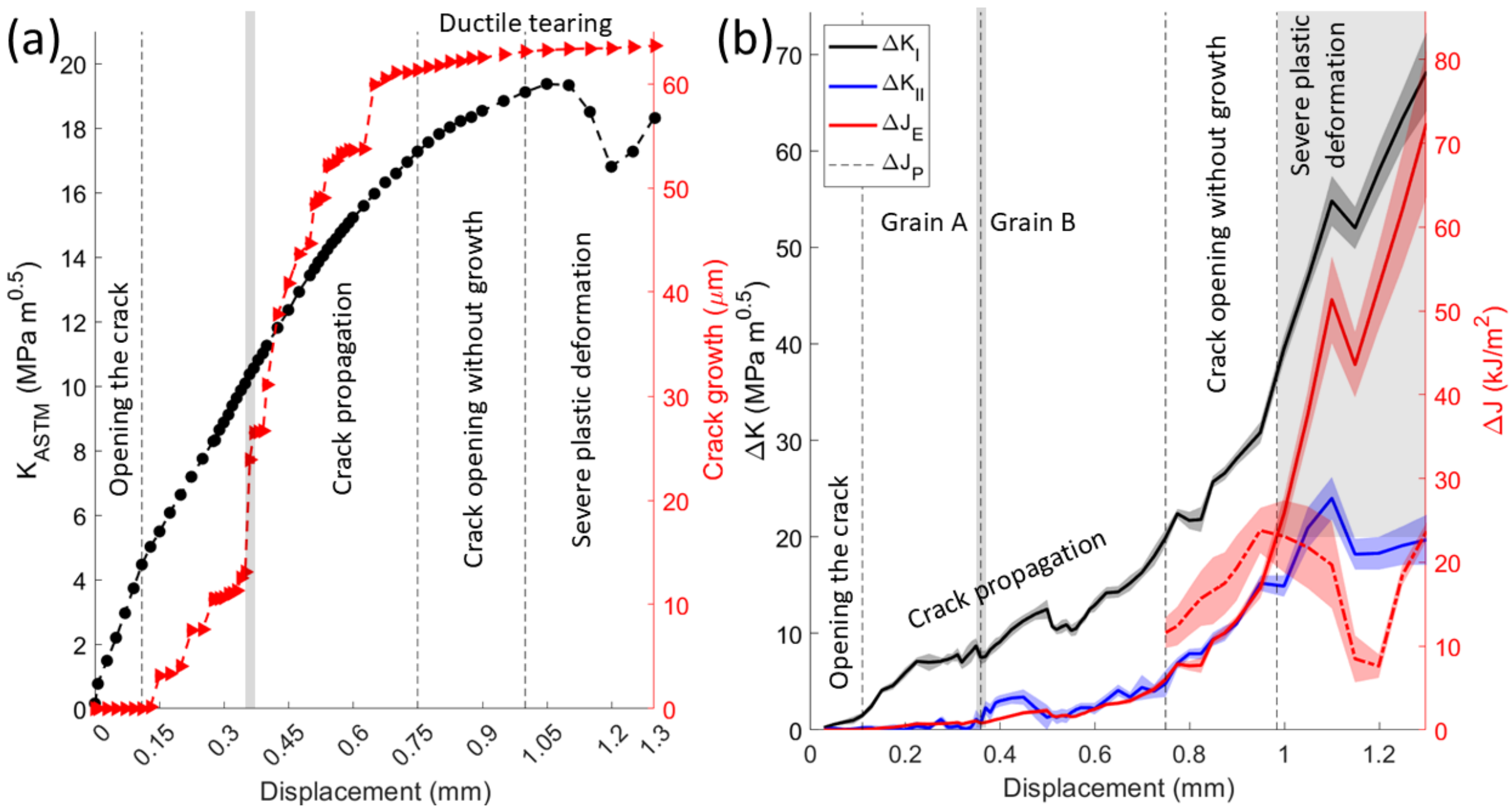

Fig. S5. The crack's stress intensity factors and strain energy release rate. (a) ASTM E1820-based stress intensity factor ($K_{ASTM}$, left axis) and corresponding crack growth (right axis) plotted against nominal displacement for the entire test. The curve is divided into key regimes: initial crack opening, crack propagation, crack opening without growth, and severe plastic deformation with ductile tearing. (b) DIC-based fracture mechanics parameters are plotted over the same displacement range. Stress intensity factors for mode I ($\Delta K_I$) and mode II ($\Delta K_{II}$), along with elastic ($\Delta J_E$) and elastoplastic ($\Delta J_P$) components of the *J*-integral, are shown. The transition from microstructure-sensitive crack propagation to macro-scale plastic deformation is indicated by the divergence of $\Delta J_E$ and $\Delta J_P$, while mode I remains dominant throughout. Shaded areas represent the region where elastic assumptions become invalid, and plasticity dominates the fracture process.

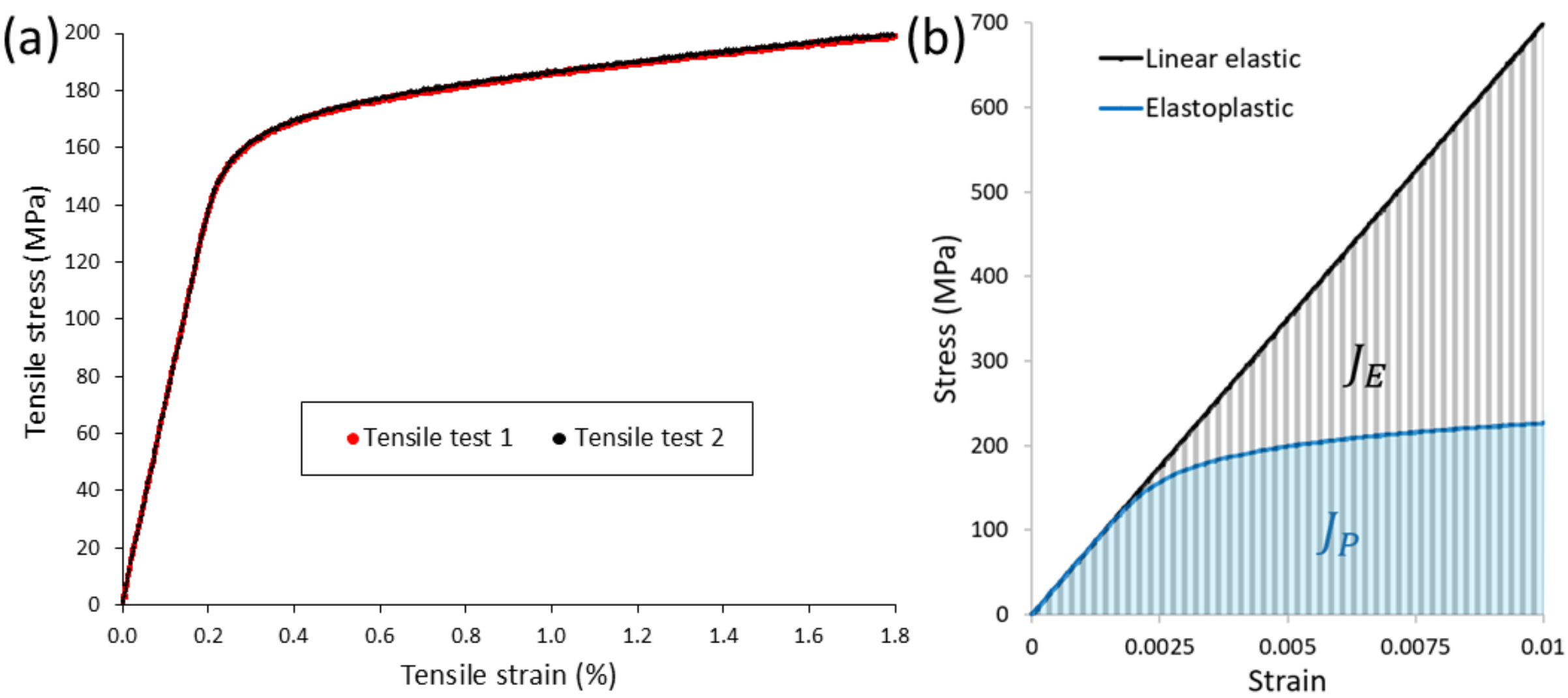

**Fig. S6. Mechanical testing of AA-5052.** (a) Tensile stress-strain curves from two independent tests on AA-5052. (b) Schematic comparison of *J*-integral estimation using linear elastic ($J_E$) versus elastoplastic ($J_P$) assumptions. The shaded regions illustrate how neglecting plasticity

can lead to overestimating the strain energy release rate when using DIC-derived displacement fields.

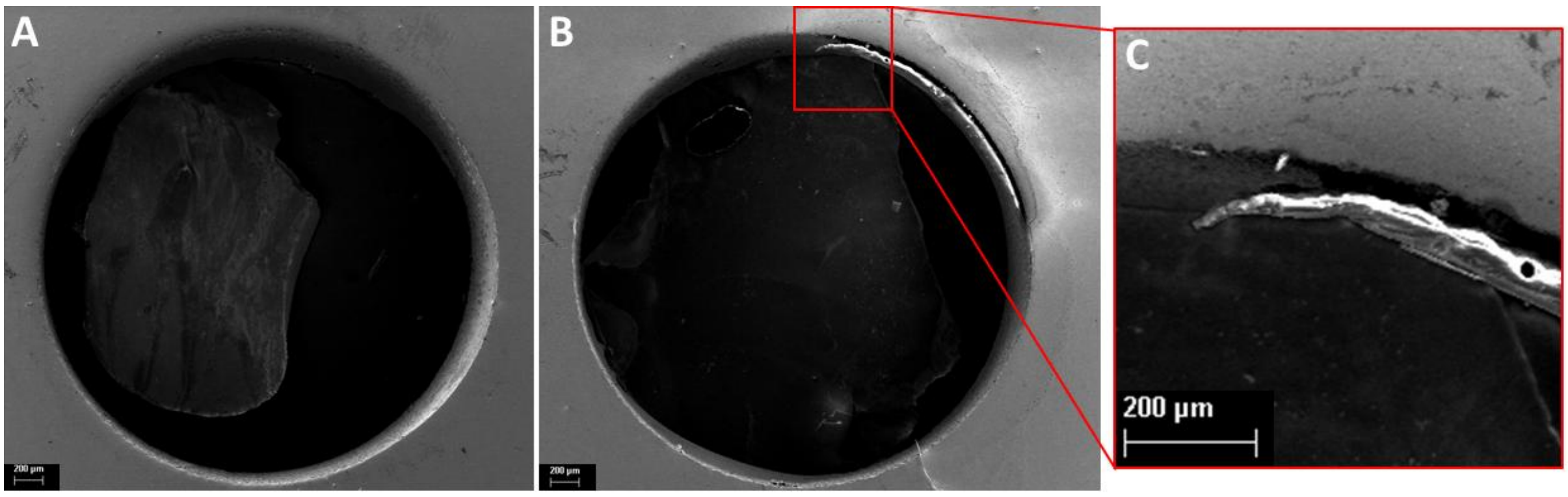

Fig. S7: The specimen grips/pinhole after the test, showing no sign of plastic damage. The focused area in C shows leftover from the carbon dag used to stick the sample on the SEM stub.

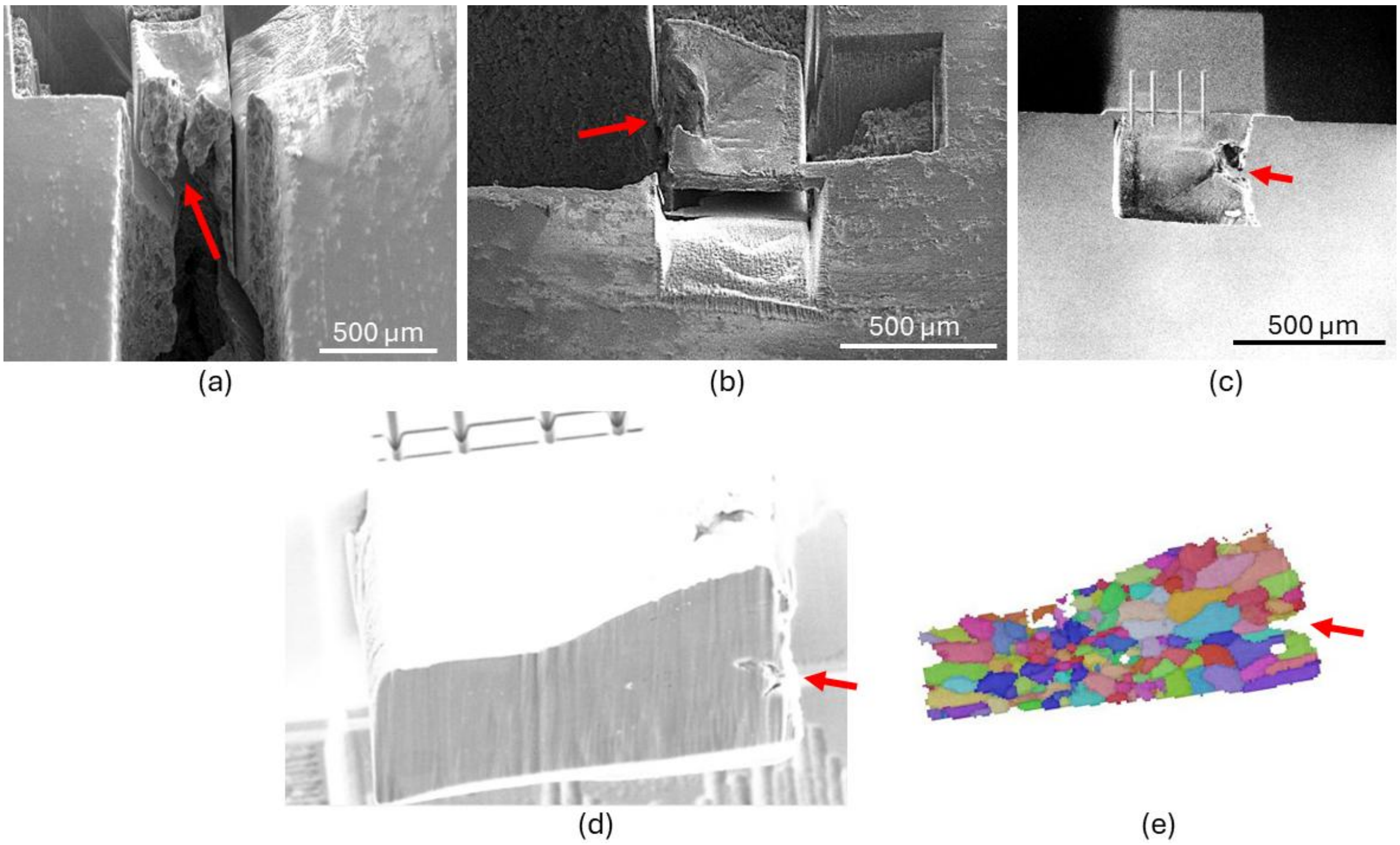

Figure S8: (a) –(b) lift out of the crack front (red arrow), (c) redeposition welding of the liftout onto the silicon wafer, (d)-(e) one of the best 3D EBSD slices.